\documentclass{article}

\usepackage{arxiv}

\usepackage[utf8]{inputenc} 
\usepackage[T1]{fontenc}    
\usepackage{hyperref}       
\usepackage{url}            
\usepackage{booktabs}       
\usepackage{amsfonts}       
\usepackage{nicefrac}       
\usepackage{microtype}      
\usepackage{lipsum}		
\usepackage{graphicx}
\usepackage{doi}

\usepackage[english]{babel}
\usepackage{amsmath}
\usepackage{relsize}
\usepackage{chngcntr}
\usepackage{nameref,hyperref}
\usepackage{amssymb}
\usepackage{subcaption}
\usepackage[verbose]{placeins}
\usepackage{graphicx}
\usepackage{algorithm} 
\usepackage{algpseudocode} 

\usepackage{makecell}

\title{Detection of Dynamical Regime Transitions with Lacunarity as a Multiscale Recurrence Quantification Measure}

\author{ \href{https://orcid.org/0000-0002-3095-8960}{\includegraphics[scale=0.06]{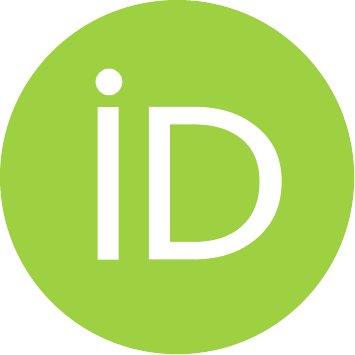}\hspace{1mm}Tobias Braun}\\
	Complexity Science (RD4)\\
	Potsdam Institute for Climate Impact Research\\
	14473 Potsdam, Germany\\
	Tel.: +49-331-28820744 \\
	\texttt{tobraun@pik-potsdam.de} \\
	\And
{\hspace{1mm}Vishnu R.~Unni} \\
	Department of Mechanical and Aerospace Engineering\\
	Princeton University\\
	NJ, USA
	\And
{\hspace{1mm}R. I.~Sujith} \\
	Indian Institute of Technology Madras\\
	Chennai 600036\\
	India
	\And
{\hspace{1mm}Juergen Kurths} \\
	Complexity Science (RD4)\\
	Potsdam Institute for Climate Impact Research\\
	14473 Potsdam, Germany\\
	\And
{\hspace{1mm}Norbert Marwan} \\
	University of Potsdam\\
	Institute of Geosciences\\
	14473 Potsdam, Germany \\
}

\renewcommand{\headeright}{ }
\renewcommand{\undertitle}{Preprint}

\hypersetup{
pdftitle={Detection of Dynamical Regime Transitions with Lacunarity as a Multiscale Recurrence Quantification Measure},
pdfsubject={q-nlin.CD},
pdfauthor={Tobias Braun, Vishnu R.~Unni, R. I.~Sujith, Juergen Kurts, Norbert Marwan},
pdfkeywords={Recurrence Plots, Regime Shifts, Lacunarity, Nonlinear time series, Thermoacoustic Instability},
}

\begin{document}
\maketitle

\begin{abstract}

We propose lacunarity as a novel recurrence quantification measure and illustrate its efficacy to detect dynamical regime transitions which are exhibited by many complex real-world systems. We carry out a recurrence plot based analysis for different paradigmatic systems and nonlinear empirical data in order to demonstrate the ability of our method to detect dynamical transitions ranging across different temporal scales.
It succeeds to distinguish states of varying dynamical complexity in the presence of noise and non-stationarity, even when the time series is of short length. 
In contrast to traditional recurrence quantifiers, no specification of minimal line lengths is required and rather geometric features beyond linear structures in the recurrence plot can be accounted for. This makes lacunarity more broadly applicable as a recurrence quantification measure.
Lacunarity is usually interpreted as a measure of heterogeneity or translational invariance of an arbitrary spatial pattern. In application to recurrence plots, it quantifies the degree of heterogenity in the temporal recurrence patterns at all relevant time scales. We demonstrate the potential of the proposed method when applied to empirical data, namely time series of acoustic pressure fluctuations from a turbulent combustor. Recurrence lacunarity captures both the rich variability in dynamical complexity of acoustic pressure fluctuations and shifting time scales encoded in the recurrence plots. Furthermore, it contributes to a better distinction between stable operation and near blowout states of combustors.
\end{abstract}


\section{Introduction}
\label{sec1}
Many efforts in nonlinear time series analysis have been dedicated to the challenge of detecting transitions between different dynamical states of a system \cite{michael2005applied, bradley2015nonlinear, scheffer2009early}. 
A broad range of real--world systems undergo such shifts between distinct regimes and their identification  provides a better understanding of the complex dynamics under study \cite{marwan2008significance, donges2015non, donges2011identification, malik2012dynamical, smirnov2017regime, malik2012dynamical, chen2016heterogeneous, unni2015multifractal,godavarthi2017recurrence}. The universality of transitions between different dynamical states for a broad spectrum of different systems elucidates why applications have been widely dispersed among many disciplines. For instance, time series in earth sciences usually require sophisticated approaches to determine abrupt changes in the complex dynamics \cite{donges2015non, marwan2018regime, goswami2018abrupt}. Regime shift detection has also gained popularity in analysis of EEG data \cite{marwan2002recurrence}, neuroscientific time series \cite{prado2014synchronization, izhikevich2007dynamical} and other medical research fields \cite{venegas2005self} where the identification of pathological regimes is crucial. Due to their complexity, financial and social time series offer interesting applications as well \cite{mantegna1999introduction, fabretti2005recurrence,malik2014fluctuation}.

Major challenges in detecing regime shifts in real-world data are often data related, e.g. by means of unevenly sampled \cite{marwan2018regime}, nonstationarity, noisy or short time series. In convenient cases, transitions are visible to the eye but usually, the exact localization of the occuring dynamic transition and, in particular, the identification of precursors poses a challenge. Segments of a time series may appear qualitatively similar at first glance but could turn out to show significantly different dynamical features. With respect to climate systems, multiple spatial and temporal scales can also hamper clear distinctions between variations in complexity of a time series. Even though regime transitions occur in a broad class of systems, data related peculiarities raise the need of a comprehensive box of tools rather than a single universal method.

In contrast to linear methods such as autocorrelation or power spectrum analysis, nonlinear techniques are able to uncover more subtle transitions in complex time series data. Multiple different approaches constitute the state-of-the-arts toolbox, ranging from complex networks \cite{krishnan2019emergence}, entropies \cite{bradley2015nonlinear}, detrended fluctuation analysis \cite{kantelhardt2002multifractal} or symbolic dynamics \cite{letellier2006estimating}. Since many empirical time series are univariate and no prior knowledge is accessible about the true dimensionality of the system, phase space reconstruction is a powerful approach to study the system's dynamics \cite{bradley2015nonlinear}. Approaches based on the phase space trajectory are closely related to the well-known Lyapunov exponents and have proven to be effective in classifying different dynamical states
\cite{malik2012dynamical, malik2014fluctuation}.

Another technique with relatively low numerial effort is the analysis of nonlinear time series by \textit{recurrence plots} (RPs) \cite{eckmann1995recurrence}. The basic idea behind this method relates back to the perspective that dynamical systems recur to states they have visited before \cite{poincare1890probleme}. As a representation of such recurrences, the binary recurrence matrix $\mathbf{R}_{ij}$ of a phase space trajectory $\mathbf{x}_i \in \mathbb{R}^d$ indicates times where the system recurs to formerly visited states by $1$s and all other times by $0$s.
It is widely used as a graphical tool but also allows for quantification of various dynamical aspects of the system under study. Since its first conception the method was successfully extended and applied to various real-world systems \cite{marwan2008historical}. The detection of regime transitions has become a prototypical field of application for RPs since it enables us to analyse complex temporal patterns of nonlinear time series in a simplified fashion \cite{marwan2007recurrence}. The majority of measures in recurrence quantification analysis (RQA) are based on black or white line structures in an RP. For instance, diagonal lines resemble parallel segments of the phase space trajectory and thus entail a degree of predictability. Approaches to capture more complex features in RPs beyond the reductionist approach of measuring line lengths have been conceived \cite{corso2018quantifying, chen2016heterogeneous}, e.g. by allowing for the entirety of possible permutations of recurrences in small submatrices. This leads to the observation that the variety of distinct microstates in deterministic and stochastic systems occupies only a fraction of the possible permutations, yielding lower entropy values. Even though this technique generally shows good robustness and does not require specification of minimum line lengths, such an approach is limited to the information captured by small submatrices on a restricted local  level. We propose a method where recurrences are also evaluated regardless of their exact orientation in the recurrence matrix, but with a surplus quantification of the scaling from the smallest to the largest possible submatrices. This can be achieved by not analysing the permutations of recurrences but by the statistics of their locally determined count.

To this extent, we make use of a measure called \textit{lacunarity} \cite{mandelbrot1983fractal, plotnick1996lacunarity}. Traditionally, it has been applied to quantify complex spatial patterns. Perhaps, the clearest interpretation of lacunarity is that it quantifies the degree of heterogeneity of the studied pattern. Often, it is applied in the context of distinguishing fractal patterns \cite{cheng1997multifractal} because objects of same fractal dimensionality can still exhibit different degrees of heterogeneity. As a rule of thumb, patterns with larger gaps yield higher values of lacunarity. Beyond this straight-forward interpretation, it classifies patterns with respect to their deviation from translational invariance \cite{cheng1997multifractal}. This highlights its applicability as a measure of heterogenity. Characterizing the heterogenity of RPs on different temporal scales using lacunarity yields meaningful information about the complexity of the underlying time series. Besides, heterogenity has already been considered as a quantifier to analyse recurrence networks \cite{jacob2017measure}.
Lacunarity has succesfully been applied to various systems ranging from characterizations of the scaling properties of the Amazon rainforest \cite{malhi2008analysis} and urban areas \cite{gomides2018lacunarity} to heterogenous patterns in bone structures \cite{marwan2007measures} and stellar mass distributions \cite{gaite2018fractal}. It is also a popular tool in Neuroscience \cite{karperien2011reviewing}. To our best knowledge, it has not yet been applied to RPs. Combining both approaches as a powerful tool to detect dynamical regime shifts is the main contribution of this work. In order to demonstrate the scope of the developed methodology, we showcase applications to both paradigmatic systems and nonlinear empirical time series.

This work is organized as follows: in Sect. \ref{sec2} we introduce our methodology by briefly summarizing the RP technique and describing the computation of lacunarity by a box-counting algorithm. In this context, we give a brief dynamical interpretation of our method. Subsequently, we study results for synthetic data from three paradigmatic systems in Sect. \ref{sec3}, namely the Logistic Map, the Roessler system and a bistable noise-driven system. The robustness of our method against noise and short time series length is examined. Finally, we provide first evidence that the proposed method is capable of detecting regime transitions in complex empirical time series. In Sect. \ref{sec4}, we identify known dynamical states in time series of acoustic pressure in a turbulent combustor \cite{tony2015detecting} and attempt to illustrate the distinction between two dynamically similar but practically contrary regimes. We conclude our findings in Sect. \ref{sec5}.
\section{Methodology}
\label{sec2}
First, we introduce recurrence plots in Sect. \ref{sec2.1} and briefly revise traditional recurrence quantification analysis. Afterwards, we define lacunarity and present detailed information for its application to RPs in Sect. \ref{sec2.2}. To gain a more profound understanding of the proposed method, we discuss its dynamical interpretation for the phase space trajectory.
\subsection{Recurrence Analysis}
\label{sec2.1}
Many real-world systems show a tendency to recur to states they have visited before. Such information can be captured by a two-dimensional visualization that may yield striking patterns which resemble the recurrences at all time instances of the time series. By studying regularities in such recurrence patterns, rich information can be obtained on the dynamics of the underlying system that go beyond the scope of linear statistical methods of time series analysis such as autocorrelation functions. In particular, different quantification measures of the visual representation prove powerful in classifying differing systems, detecting non-linear correlations and identifying dynamical regime transitions. The basic concept can be outlined by defining the recurrence matrix
\begin{align}
\mathbf{R}_{ij} \, = \, \left\{
\begin{array}{ll}
1 & \mathrm{if} \  ||\mathbf{x}_i - \mathbf{x}_j|| \leq \epsilon \\
0 & \mathrm{if} \ ||\mathbf{x}_i - \mathbf{x}_j|| > \epsilon \\
\end{array}
\right.
\end{align}
\label{eq1}
with a time series $\mathbf{x}$ at two arbitrary times $i$ and $j$ and a suitable norm $\|\cdot\|\,$. The vicinity threshold $\epsilon$ needs to be fixed with respect to the distances of time series values such that a meaningful expression of recurrences is obtained. A popular approach to do so is, for example, to choose a value which entails a fixed recurrence rate for the RP \cite{marwan2007recurrence}. The resulting visual representation of a recurrence matrix is a binary image of black and white dots from which temporal patterns can be inspected. For higher dimensional systems, recurrence analysis needs to be based on a phase space representation of the time series as a trajectory in $d\,$-dimensional space. Despite the fact that the dimension of the regarded system is often unknown, Takens' theorem \cite{takens1981detecting} ensures that a time--delay embedding can be found that gives an appropriate phase space reconstruction. To this extent, an embedding delay is often specified prior to fixing the optimal embedding dimension for the system. We will apply the broadly used mutual information criterion and the False Nearest Neighbours (FNN) method \cite{cao1997practical, pecora2007unified} to estimate both parameters.

For many systems, RPs have been successfully applied to reveal a high degree of complexity both in terms of nonlinear dynamics and stochastic fluctuations \cite{marwan2007recurrence}.
In the detection of regime transitions, a reliable quantifier that yields an unambiguous distinction of dynamical regimes is generally required. Traditional complexity indicators such as Lyapunov Exponents \cite{kantz1994robust} are not always robust against noise and require rather long time series. They are also often not able to capture the relevant time scales of the system's shifting dynamics which is particularly important for systems with multiple characteristic time scales. Yet, such information is contained in RPs and can be uncovered using recurrence quantification measures \cite{webber2015recurrence}. Most of them are based on the statistical distribution of line structures in RPs. For instance, diagonal lines of certain length indicate a similar evolution of different segments of a time series. A popular quantification of an RP based on diagonal lines is defined as the fraction of lines that exceed some specified minimal line length. As it quantifies the degree of determinism in a time series, it is refered to as $\mathrm{DET}$.
Vertical lines indicate that the system is trapped in a certain phase space region for subsequent times. White vertical gaps indicate transitions between different phase space regions while black square-like structures point at time intervals in which the system remains confined in a small region of the phase space. If such patterns reoccur with statistical significance, they uncover regularities of the time series and may yield characteristic recurrence time scales. Line--based recurrence measures have been applied to a diversity of complex real--world systems as complexity measures to uncover transitions \cite{marwan2007recurrence}. Yet, different embedding parameters can yield varying results and edge effects as well as high sampling rates might result in spurious quantifications \cite{marwan2011avoid}, thus requiring corrections \cite{kraemer2019border}. On top of that, the application of line-based RQA measures is limited to systems that do not show more complex recurring patterns which are refered to as microstates of an RP \cite{corso2018quantifying}. Related complexity measures that go beyond this scope have shown that they can have superior performance \cite{malik2014fluctuation, chen2016heterogeneous}. In this work, will put forward a novel RQA measure that characterizes the heterogenity of an RP and is not based on certain microstructures.
\subsection{Recurrence Lacunarity}
\label{sec2.2}
Lacunarity is often illustrated as a measure of `gappiness' or as a property that can characterize the heterogenity of a spatial pattern. It was introduced to distinguish between different fractal patterns of equivalent fractal dimensionality. However, its scope goes beyond the distinction of fractal structures. More formally, we can regard it as a measure of deviance of a pattern from translational homogenity \cite{plotnick1996lacunarity, quan2014lacunarity}. As lacunarity is usually derived for different spatial scales, it is possible to identify a certain scale above which a pattern is translational invariant and below which it is too heterogenous to be regarded as such.

Methods to calculate lacunarity for empirical patterns are often based on box--counting algorithms \cite{karperien2016box} similar to those used in estimating fractal dimensionality of a pattern. Essentially, a grid is applied to the studied pattern and pixels in each box are counted. The specific algorithm to compute lacunarity for RPs is summarized in algorithmic form below.
Even though box-counting algorithms can also be modified so that an analysis of grayscale- \cite{dong2000test} or RGB--encoded \cite{ivanovici2009fractal} patterns is possible, in application to RPs a basic algorithm for binary patterns suffices. The quantity that enables us to analyse scaling properties of a complex pattern is the size of boxes on the applied grid.
Given an RP as a $T\times T\,$--matrix and a fixed box size $w$, the mass $M$ of each box is obtained by counting the black pixels inside the box. This results in a mass distribution $P_w(M)$ for all $N$ boxes. From this distribution, we compute the moments
\begin{align}
Z^{(q)}(w) \ = \ \sum_M M^q P_w(M) \ .
\end{align}
\label{eq2}
In the definition of lacunarity $\Lambda$, only the first and the second moments $Z^{(1,2)}$ are considered:
\begin{align}
\tilde{\Lambda}(w) \, = \, \frac{Z^{(2)}(w)}{\left[Z^{(1)}(w)\right]^2} \, = \, 1 + \frac{\sigma^2(w)}{\mu^2(w)}
\end{align}
\label{eq3}
with mean $\mu$ and standard deviation $\sigma$.
To have a measure in $[0,1]$, we normalize lacunarity by also computing the lacunarity $\tilde{\Lambda}^{\dagger}$ of the complement of the set ($1$s replaced by $0$s and \textit{vice versa}). Consequently, we define it as
\begin{align}
\Lambda(w) \ = \ 2 - \left( \frac{1}{\tilde{\Lambda}(w)} + \frac{1}{\tilde{\Lambda}^{\dagger}(w)} \right) \ .
\end{align}
\label{eq4}
This refined definition of lacunarity also enhances the detection of significant gap sizes compared to eq. (3) and is thus prefered \cite{marwan2007measures}. Various box--counting methods beyond the basic approach employed in this work are known (e.g. gliding box-counting \cite{allain1991characterizing}) and generalized versions exist (e.g. for multifractal data \cite{alber1998improved}). The often used standard gliding box approach results in a higher number of boxes but is also known to cause biased values due to edge effects \cite{feagin2007edge}. Note that in any case, minimum and maximum box size have to be chosen based on the time series length.

\begin{algorithm}
	\caption{Recurrence Lacunarity} 
	\begin{algorithmic}[1]
	\If{embedding required}
		\State Choose $d,\, \tau$ and embedd time series
	\EndIf
	\State Compute RP $\underline{\underline{R}}$ with specified vicinity threshold $\epsilon$
		\For {box width$\, w\,=2,3\ldots \,\ll T$}
			\State Split $(T\times T)$-matrix $\underline{\underline{R}}$ into $N = \lfloor T/w \rfloor
			\times \lfloor T/w \rfloor$ disjunct boxes
			\For {each box $b_w^{(i,j)}$}
				\State Count recurrences: $M = \sum_{i',j'}^{i,j}\delta(\underline{\underline{R}}_{i'j'}-1)$	
			\EndFor
			\State Compute normalized lacunarity $\Lambda$ for fixed $w$ 		            from eq. (2-4)
			\State Apply \emph{bootstrap} by randomly drawing sufficient number of boxes $b_w^{(i,j)}$ for significance testing
			\If{$T\%w \neq 0$}
				\Repeat \, iteration with varied grid position
				\Until robustness ensured
			\EndIf
		\EndFor
	\end{algorithmic} 
\end{algorithm}
\FloatBarrier

In the application of lacunarity to RPs, we refer to it as \textit{recurrence lacunarity} (RL).
In view of RPs, black pixels are equivalent to recurrences of a trajectory in reconstructed phase space. Thus, we are effectively carrying out an analysis of local (in a temporal sense) recurrence statistics and quantification of local variations. This is essentially implemented via the computation of variance of recurrence points contained in boxes which are located at different positions in the RP. An extension to higher statistical moments is also conceivable \cite{valous2018multilacunarity}.
Our approach circumvents the necessity of defining any sort of microstate and is not restricted to the usual statistical analysis of line structures in the RP. The scaling (sucessive increasing of time intervals) is expected to pinpoint  relevant temporal scales related to average recurrence times and quasi--periodicities. This is confirmed by the displayed recurrence lacunarity curves in Fig. \ref{fig2} from which some will be studied in more detail in Sect. \ref{sec3}.
\begin{figure*}[!h]
\begin{subfigure}[t]{.32\textwidth}
\centering
\includegraphics[width=1\textwidth]{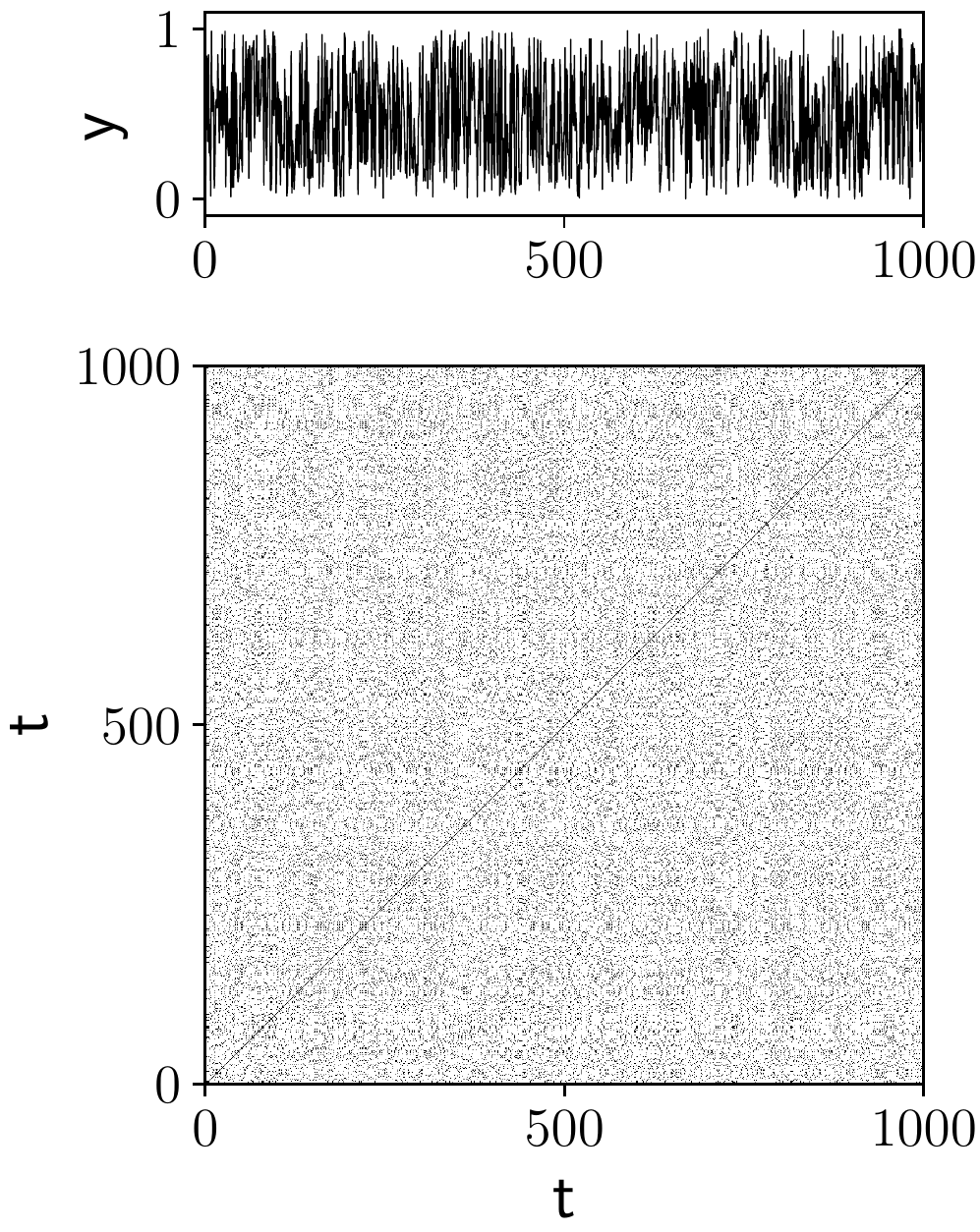}
\label{fig1a}
\caption{White noise}
\end{subfigure}
\begin{subfigure}[t]{.32\textwidth}
\includegraphics[width=1\textwidth]{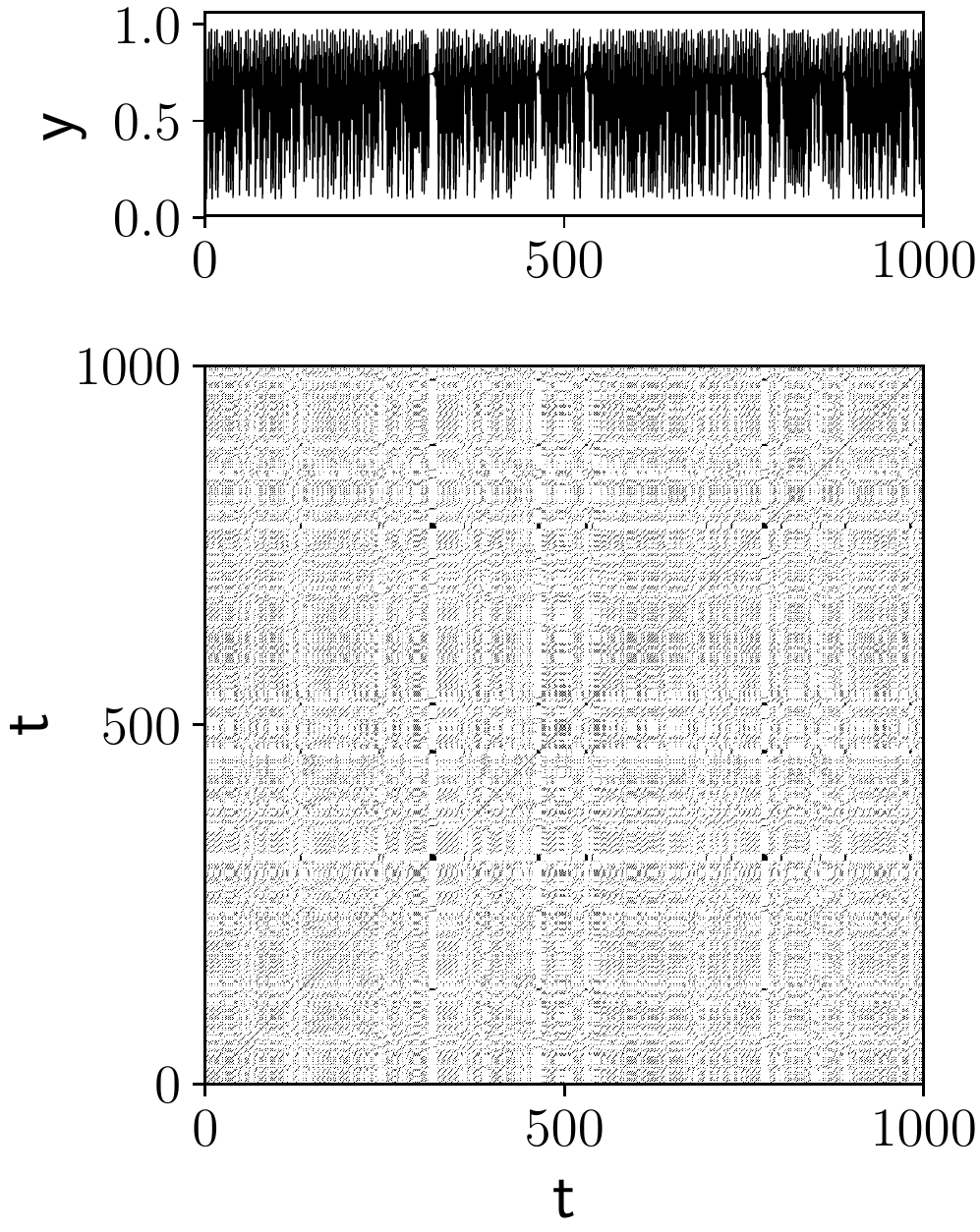}
\label{fig1b}
\caption{Logistic Map \\ ($r=3.9$)}
\end{subfigure}
\begin{subfigure}[t]{.32\textwidth}
\includegraphics[width=1\textwidth]{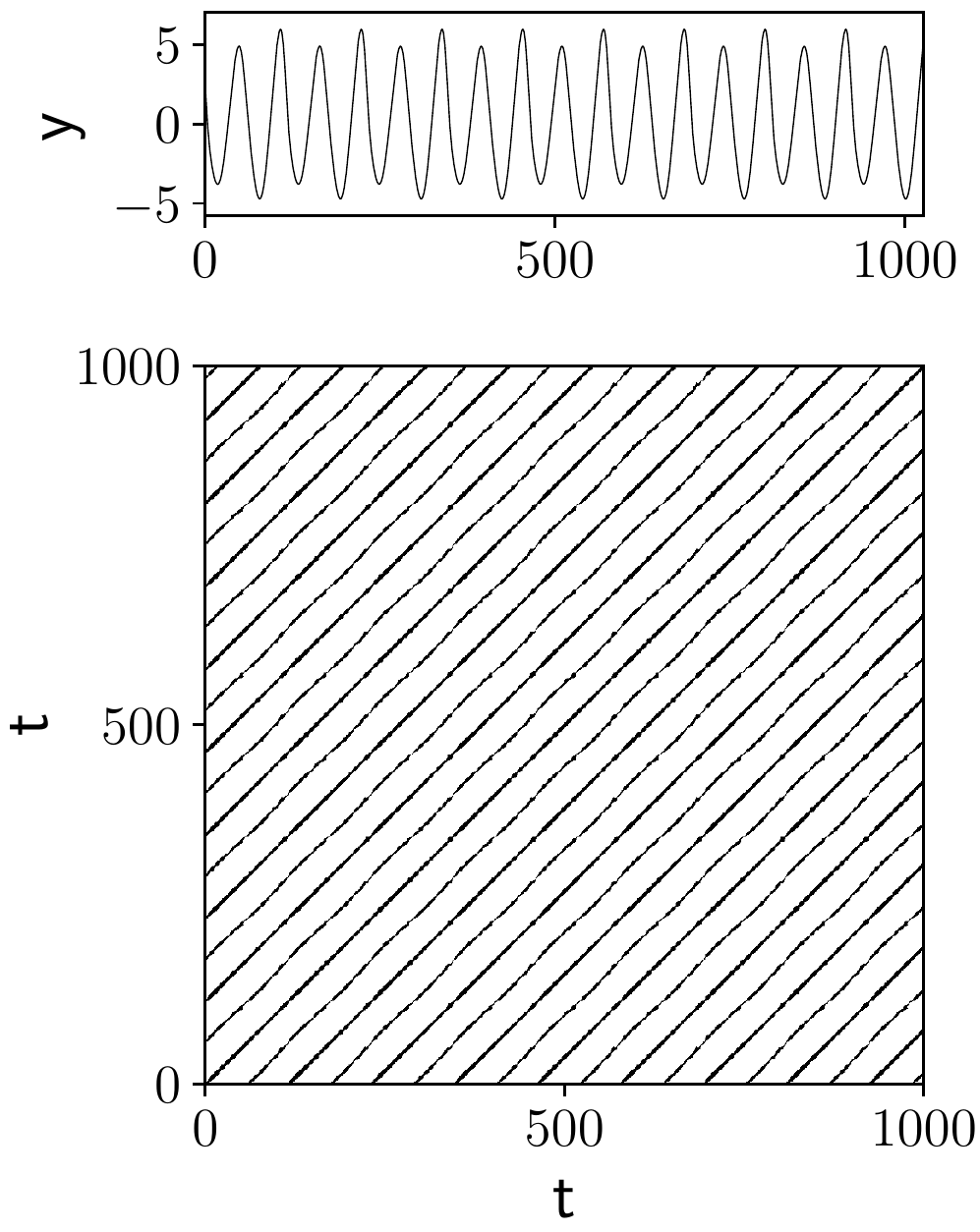}
\label{fig1c}
\caption{Roessler system (periodic, $a = 0.2,\ b = 0.2,\ c = 3$)}
\end{subfigure}
\begin{subfigure}[t]{.32\textwidth}
\includegraphics[width=1\textwidth]{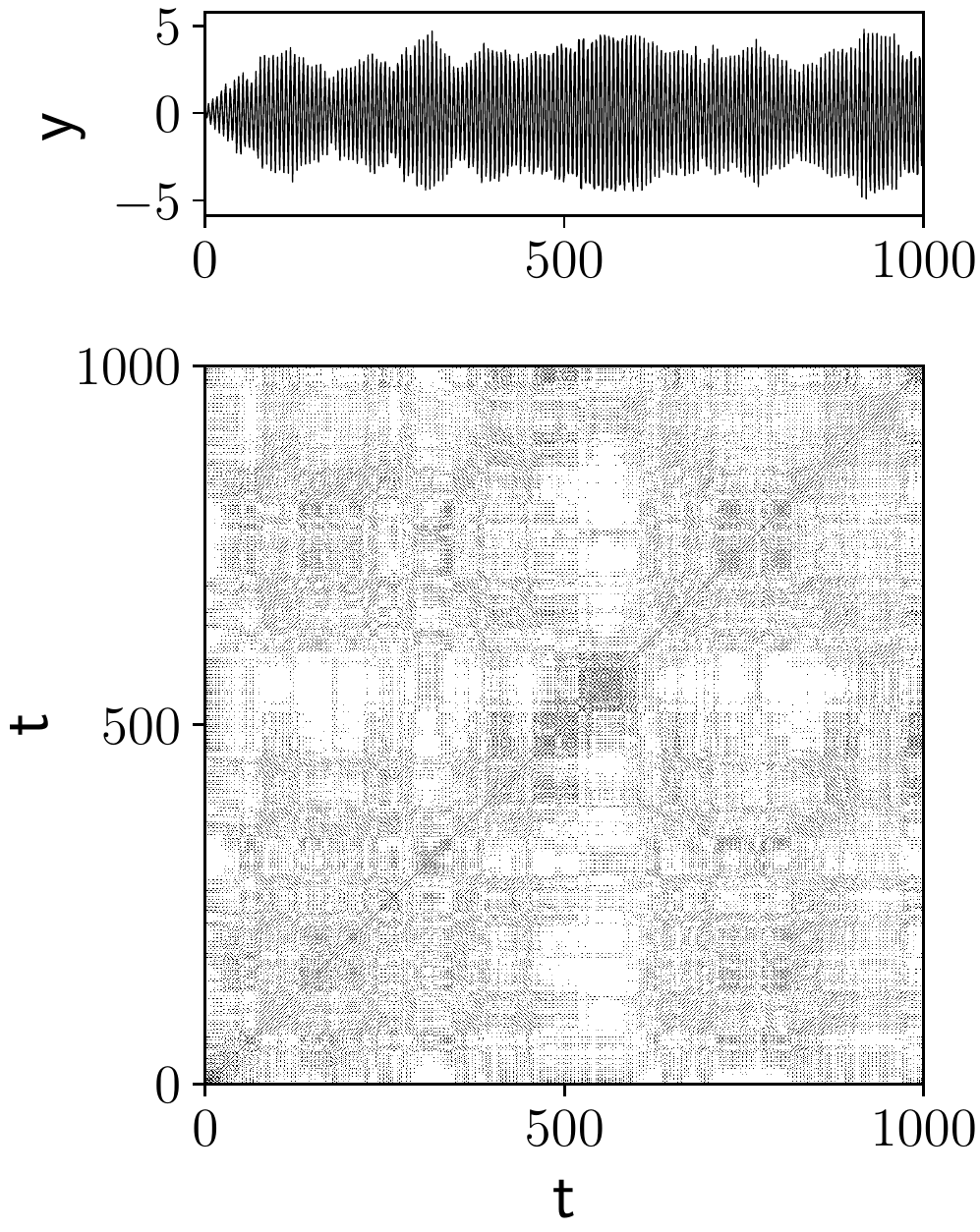}
\label{fig1d}
\caption{AR(2)--process \\ ($\mu_1 = 1,\ \mu_2 = -1,\ \theta = 0.5$)}
\end{subfigure}
\begin{subfigure}[t]{.32\textwidth}
\includegraphics[width=1\textwidth]{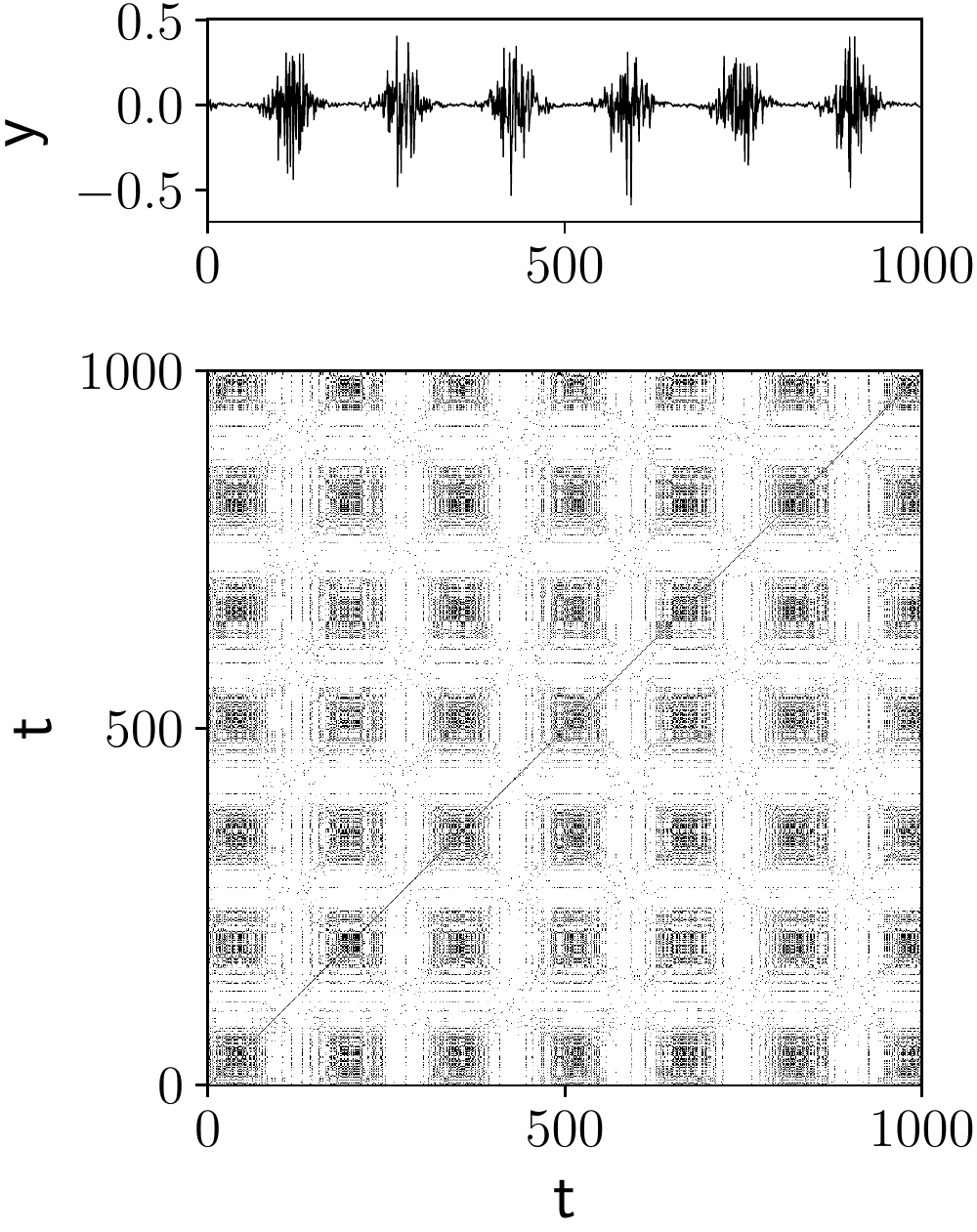}
\label{fig1e}
\caption{MFGN--process with sinusoidally tuned Hurst Exponent between $[0.25, 0.75]$}
\end{subfigure}
\begin{subfigure}[t]{.32\textwidth}
\includegraphics[width=1\textwidth]{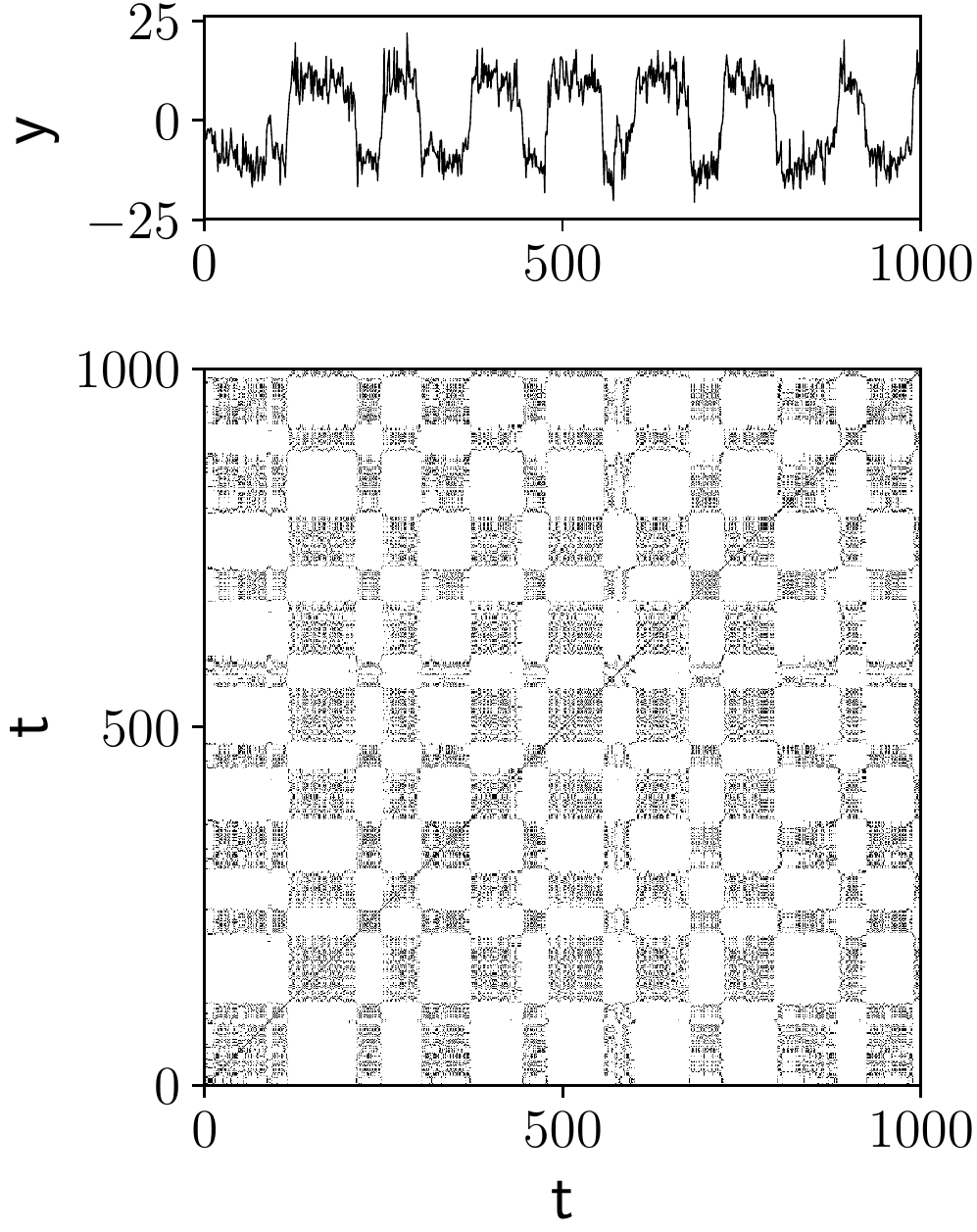}
\label{fig1f}
\caption{Bistable white noise-driven system ($K=100, \,A=300$, $\,\omega=10, \,D=45$)}
\end{subfigure}
\caption{Recurrence plots for time series of different deterministic and stochastic systems.}
\label{fig1}
\end{figure*}
\begin{figure}
\centering
\includegraphics[width=.5\textwidth]{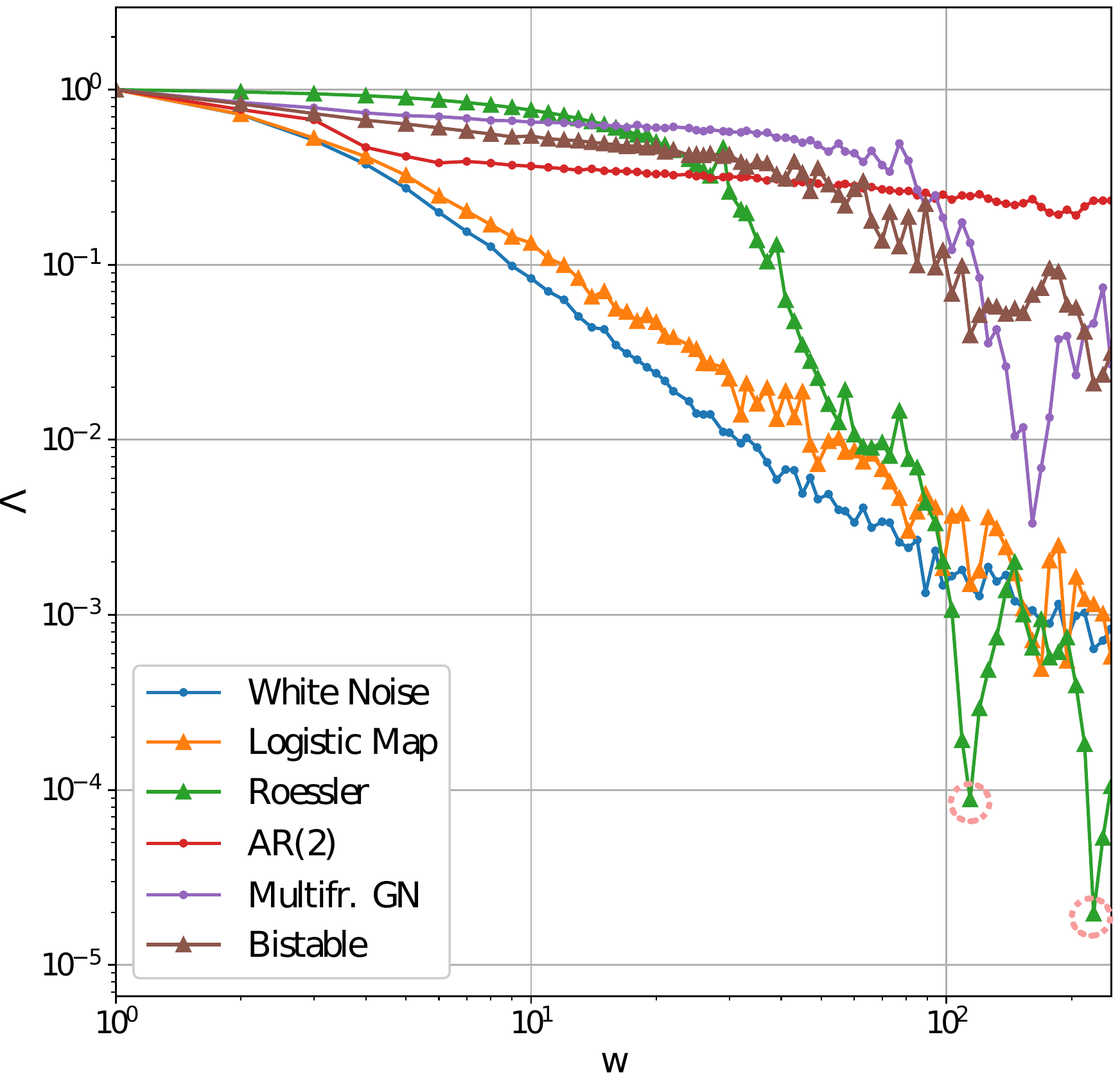}
\caption{Recurrence lacunarity curves $\Lambda(w)$ in double-logarithmic plot, corresponding to displayed RPs in Fig. \ref{fig1}.}
\label{fig2}
\end{figure}

Figure \ref{fig1} shows examples of RPs of systems that show fundamentally different dynamics. While a white noise process entails an RP with randomly distributed black dots, a Logistic Map in the chaotic regime ($r = 3.9$) still results in some local structures such that RL is higher for almost all $w$, resembling higher complexity. A Roessler system in the periodic regime generates well pronounced lines corresponding to deterministic periodic dynamics. As expected, the characteristic width between the lines is captured in the variation of RL as a local minimum (left dotted circle) since for boxes of the same size, homogenity is enhanced. The second visible minimum (right dotted circle) is located at twice the period of the time series. Stochastic signals such as an AR(2)--process and Multifractal Gaussian Noise (MFGN) \cite{bacry2001multifractal} yield a higher degree of complexity in terms of low translational invariance of the corresponding RPs. Both the MFGN and the bistable noise-driven system (see sec. \ref{sec3.3}) time series have a visible periodic modulation which is resembled by the distinct gap sizes in the respective RPs. For the former time series, RL sharply drops when $w$ reaches the gap size. For the latter, the stochastic component results in slight variations of the gap size but still, RL captures them as a local minimum at $w\approx 10^2$.

\subsubsection{Dynamical Interpretation}
\label{sec2.2.1}
\begin{figure*}[!h]
\centering
\includegraphics[width=.83\textwidth]{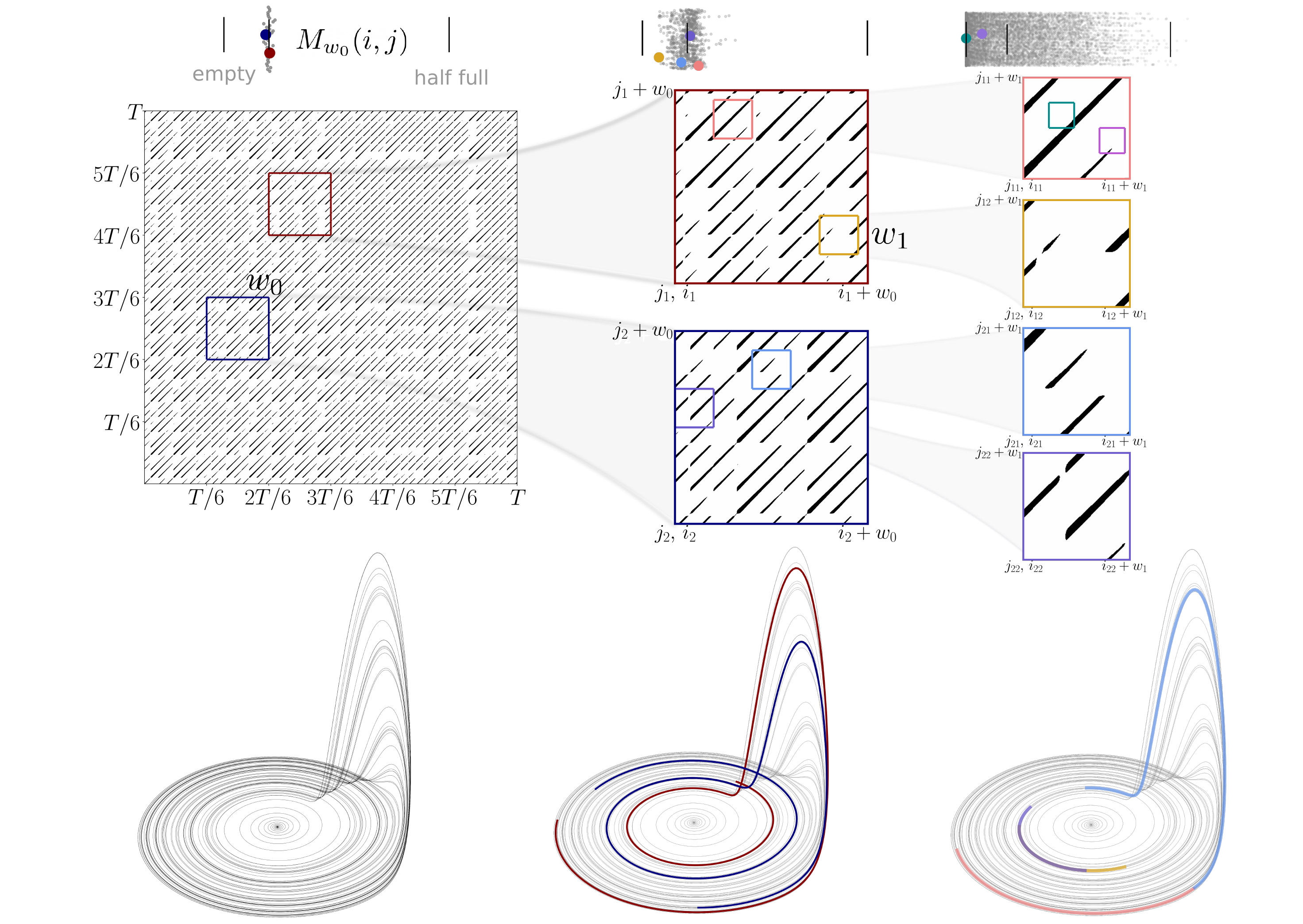}
\caption{Schematic illustration of the relation between box-counting on RPs and the underlying phase space trajectory for the Roessler attractor  ($a=0.1,\, b=0.1,\, c= 14$). The full RP is displayed on the left with two exemplary boxes (blue and red). Above, the box counts $M_{w_0}(i,j)$ of all boxes (although not indicated) located at grid positions $(i,\,j)$ are shown with the average box count (vertical line) whereas their index is chosen as a y-coordinate for better visibility. From left to right, boxes of decreasing width $w$ are zoomed in and additional boxes are indicated. The segments of the phase space trajectory that correspond to the boxes are color coded respectively. The scatter plots illustrate that for decreasing box widths, heterogeneity by means of dispersion of the box counts increases.}
\label{fig3}
\end{figure*}

As standard RQA measures are based on line structures in the RP, they have clear interpretations in terms of the phase space trajectory and their relationships to dynamical invariants (such as Lyapunov exponents and Renyi entropy) are well established \cite{marwan2007recurrence}. These questions also arise for RL. 
Picking a box within an RP and deriving some related statistics can be interpreted as sampling two segments of a phase space trajectory. If we denote the starting point of the first trajectory by $i$ and the second by $j$, the lower left corner of the box is located at $(i,j)$ and its upper right corner at $(i+w,j+w)$. The number $M_{i,j}$ of recurrences contained in the box characterizes the similarity between the two trajectory segments by means of their recurrences. For fixed $\epsilon$, these are equivalent to the number of contained phase space vectors that are nearest neighbours by means of a low distance in phase space.
Figure \ref{fig3} illustrates the relation between box-counting of recurrences and the phase space trajectory based on the Roessler attractor as a paradigmatic example (see Sect. \ref{sec3.2}
). RL quantifies the heterogeneity of recurrent temporal patterns representing different segments of the phase space trajectory.
For increasing length of trajectory segments (from right to left in Fig. \ref{fig3}), their recurrence as well as their divergence can be captured by means of recurrence patterns encoded in the RP. 

Apart from the dynamical interpretation, it is an open question how fractality of RPs is related to self-similarity in the underlying time series or the attractor \cite{babinec2005global}. As a general conception, the very basic structures of RPs such as diagonal lines, vertical lines and blocks show some resemblance to typical 1- and 2-dimensional fractals like the Cantor set or the Sierpinski carpet \cite{lin1986suggested}. Self--similarity by definition arises from the spatial recurrence of patterns \cite{webber2012recurrence}.
Even though ideal monofractal patterns can not be expected to occur in a pure fashion in RPs, some degree of self--similarity appears intuitive:  small-scale recurring patterns often constitute the correlations on longer time scales which resembles what is displayed in an RP.
Self--similarity over some range of box sizes applied to an RP could thus be interpreted as a similar tendency to recur to formerly visited states regardless of the respective time scale.
On top of that, it is known that geometrical quantities computed from recurrence networks show relations to the phase space dimension and exhibit fractal scaling properties \cite{donner2011geometry}.
This ultimately raises the question whether there is a direct relation of RPs fractality to the degree of fractality of the time series and the attractor. This should be explored in more detail in a future study and is beyond the scope of this work.

\section{Dynamical Transitions in Synthetic Data}
\label{sec3}
We demonstrate the ability of RL to capture different kinds of regime shifts by its application to model data from different dynamical systems (Table 1). These systems cover some of the important aspects that need to be accounted for if transitions in real data should be identified. 
In Sect. \ref{sec3.1}, we study transitions between chaotic and periodic dynamics for the Logistic Map. In Sect. \ref{sec3.2}, transitions in the Roessler System as a three-dimensional continuous system are analysed. A bistable noise-driven stochastic process is examined in Sect. \ref{sec3.3} in order to demonstrate the ability of our approach to uncover rather subtle transitions. Finally, we examine an experimental dynamical systems in Sect. \ref{sec4}.
\begin{table*}
\begin{center}
\begin{tabular}{ l | l | p{6.5cm} | p{1cm} | p{2cm} }
\textbf{System} & \textbf{Dynamical regimes} & \textbf{Equations} & \textbf{Param.} & \textbf{Embedding} \\ \hline\hline
\makecell{\\ \\ Logistic Map} & \makecell{\\ \\ deterministic/chaotic} & \begin{equation*} x_{n+1} \, = \, rx_n(1-x_n) \qquad \  ,\  r\in[3.5,3.95]\end{equation*} & \begin{center} $r$ \end{center} & \begin{center} \textbf{/} \end{center} \\ \hline
\makecell{\\ \\ Roessler} & \makecell{\\ \\deterministic/chaotic} & \begin{equation*}
\begin{split}
\dot{x}(t) \, = \, -(y(t) + z(t))\qquad \  ,\  a=0.2 \\
\dot{y}(t) \, = \, x(t) + ay(t) \qquad\quad\ ,\  b=0.2  \\
\dot{z}(t) \, = \, b + (x(t)-c)z(t)  \quad ,\ c\in[2,10]
\end{split}
\end{equation*} & \begin{center} $c$ \end{center} &
 \begin{equation*}
\begin{split}
d = 3, \\
\quad \left. \tau\middle/\Delta t\right. = 18 
\end{split}
\end{equation*}
\\ \hline

\makecell{\\ \\ Bistable noise-driven} & \makecell{\\ \\ determ./stochastic} &
\begin{align*}
\begin{split}
\dot{x}(t) \, = \, \left[ K\left(x(t) - x^3(t)\right) + A\, \mathrm{cos}\,\omega t \right] \, + \\ \,  D \xi(t) \\ 
(K = 100,\, A = 320)
\end{split}
\end{align*}
& \begin{center} $\omega,\, D$ \end{center} & \begin{center} \textbf{/} \end{center} \\ \hline
    
\makecell{\\ \\Thermoacoustic\\ combustor} & \makecell{\\ experimental} & \begin{center} \textbf{/} \end{center} & \begin{center} equiv. ratio $\phi$ \end{center} &
\begin{equation*}
\begin{split}
\qquad d = 5, \\
\qquad \tau = 23 
\end{split}
\end{equation*} 
\\ \hline
\end{tabular}
\end{center}
\caption{Overview of studied systems for detecting dynamical regime shifts. The systems are distinguished based on their exhibited dynamical regimes, underlying equations, transition parameters and delay-embedding parameters.}
\end{table*}
\label{tab1}

RPs reveal various structures \cite{corso2018quantifying} on a broad range of time scales. RL captures these by means of variations in the characteristics of the power law scaling of RL with $w$. We will visualize this by showing the variation of the power law for each of the systems and refer to it as the \emph{RL curve}.
Moreover, we characterize the scaling of RL curves by computing the slope $\alpha$ of $\mathrm{log}\Lambda$ against $\mathrm{log}w$ by linear regression. Note that if multiple scaling regions coexist for one RP, a single slope will yield ambiguous results and other indicators should be chosen (e.g. the regression error).

Complexity measures indicate transitions by changing to significantly high or low values. In order to assess significance, we apply the following bootstrapping method that entails confidence intervals: we draw a random sample (with repetitions) from all boxes the RP was divided into and derive RL only from this sample (see Algorithm 1). This resampling procedure is repeated $N$ times to obtain a distribution of RL values that jitter around the true value. The $5\%/95\%\,$-- quantiles characterize the width of this distribution and are used as confidence bounds. Similar approches are usually employed with other complexity measures \cite{tibshirani1993introduction, lancaster2018surrogate, marwan2013recurrence}. 
In most applications, confidence bounds should reflect significance equally sufficient for all values obtained for the respective complexity measure (i.e. for each configuration of the parameters that control regime transitions). In our case, this consequently raises the question whether to sample from the global or local distribution of box counts. In the local case, box counts for a single window are used to compute confidence bounds whereas in the global case, box counts from all windows are joined. In both cases, the calculation of quantiles is performed such that only a single parameter-independent value is obtained, yielding horizontal lines that indicate the separation between `regular' dynamics and dynamical transitions.
Extensive comparative analysis of the resulting confidence bounds indicates that the latter approach yields more convincing results for the studied systems. Bootstrapping from the global box-count distribution generally yields more narrow bounds that seem to overestimate the true number of regime shifts.
Consequently, for an underlying series of length $T$, we obtain $T$ locally bootstrapped RL distributions of size $N$ from which we compute $95\%\,$-- confidence bounds.

\subsection{Logistic Map}
\label{sec3.1}
As a first paradigmatic example, we employ the Logistic Map \cite{ott2002chaos} as a system that is well-known to produce pronounced transitions between periodic, chaotic and laminar behaviour.
The variation of the parameter $r$ introduces bifurcations between regimes of different complexity as displayed in Fig. \ref{fig3}. While increasing $r$ entails a tendency of less predictability, windows of periodic dynamics arise in between. For each parameter value $r \in [3.5, 3.95]$ with $\delta r = 0.00045$ we generate a time series of suitable length such that after discarding transients, we have $T=1000$ points in time. For each such time series $x_r(t)$, we calculate an RP. We choose the threshold $\epsilon=0.1\sigma_{\mathrm{ts}}$ with $\sigma_{\mathrm{ts}}$ being the standard deviation of $x_r(t)$. From each RP, we compute RL for a number of $k=70$ different box sizes $w\in [2, T/4]$. This results in a single RL curve like in Fig. \ref{fig2}. 
A standard measure to detect regime transitions for such paradigmatic systems is the largest Lyapunov exponent $\lambda_1$ \cite{bradley2015nonlinear}. We use it as a reference for our results to evaluate the detection of transitions. 
\begin{figure}[!h]
\centering
\includegraphics[width=.65\textwidth]{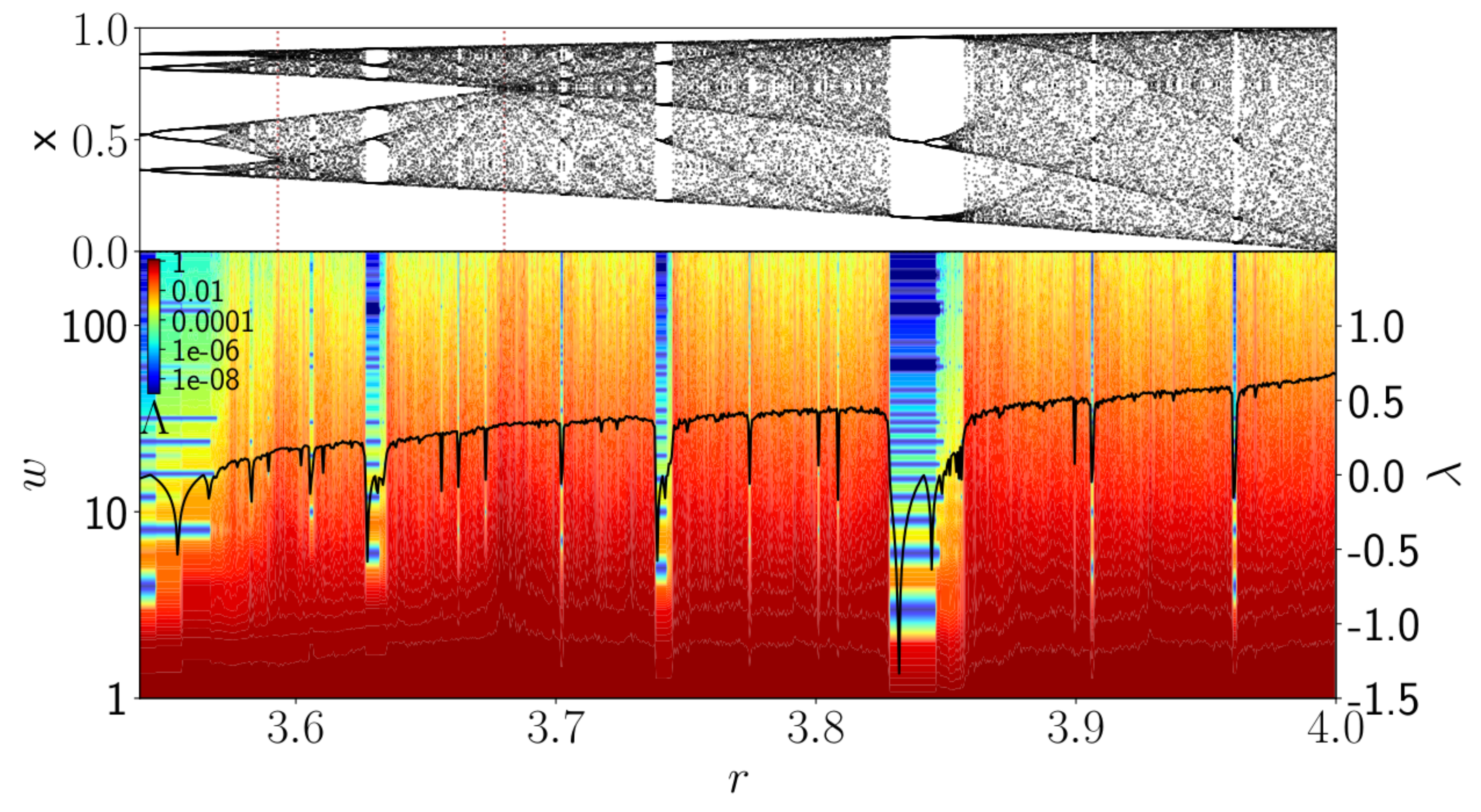}
\caption{Bifurcation diagram of Logistic Map and RL curves for varying $r \in [3.5, 3.95]$ with $n=2000$. Dashed vertical lines indicate chaos-chaos transitions. Each time series has $T=1000$ values after discarding transients. For the RL curves, $w\,$--axis and color coding are scaled logarithmically. The black curve illustrates the variation of $\lambda_1$.}
\label{fig4}
\end{figure}
\FloatBarrier
Below the bifurcation diagram in Fig. \ref{fig4}, RL is displayed using color coding. Each RL curve is plotted in double-logarithmic coordinates. Periodic windows are clearly detected as the corresponding RPs are homogenuous at all time scales. 
Furthermore, it appears that e.g. around $r\approx 3.68$ (red vertical line) it identifies a transition to a regime not well captured by $\lambda_1$ which is known to arise from the intersection of the \textit{supertrack functions} \cite{marwan2002recurrence}. 
At these, an unstable singularity results in laminar behaviour i.e. the time series becoming `trapped' in a certain range of values for some time intervals. RL is able to detect such chaos--chaos transitions.

\begin{figure}
\begin{subfigure}[t]{.51\textwidth}
\centering
\includegraphics[width=1\textwidth]{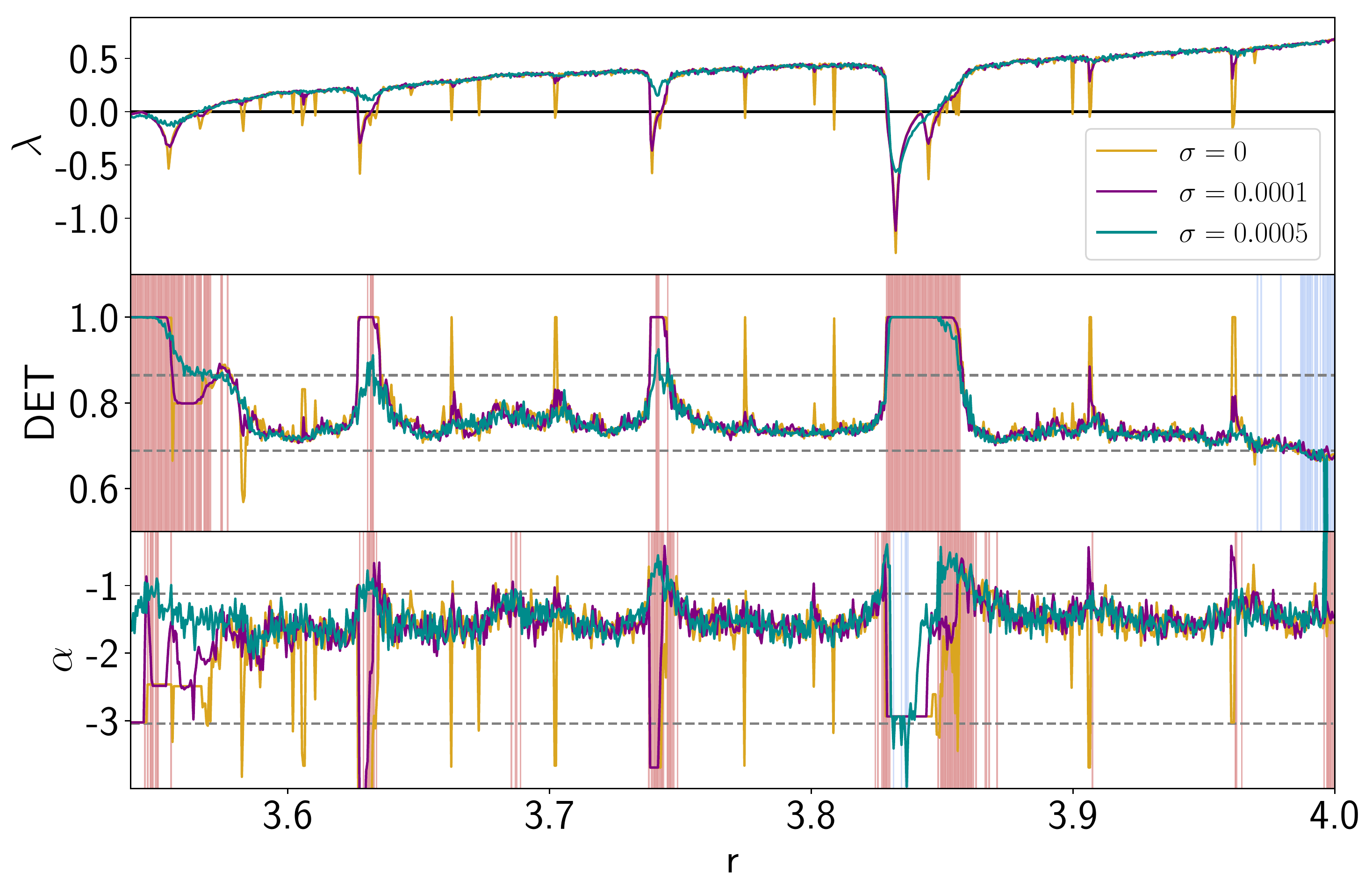}
\caption{$\lambda_1$, DET and the slope $\alpha$ of the RL curve for for varying $r \in [3.5, 3.95]$ with $n=1000$. Different curves correspond to different noise intensities. Each time series has $T=1000$ values after discarding transients. $95\%\,$-- confidence bounds (gray dashed lines) for both RQA measures are obtained via bootstrapping and refer to the noise-free case. Significant transitions outside of these are indicated by blue/red rectangles for noise with $\sigma = 0.0005$ .}
\label{fig5a}
\end{subfigure}
\begin{subfigure}[t]{.49\textwidth}
\centering
\includegraphics[width=1\textwidth]{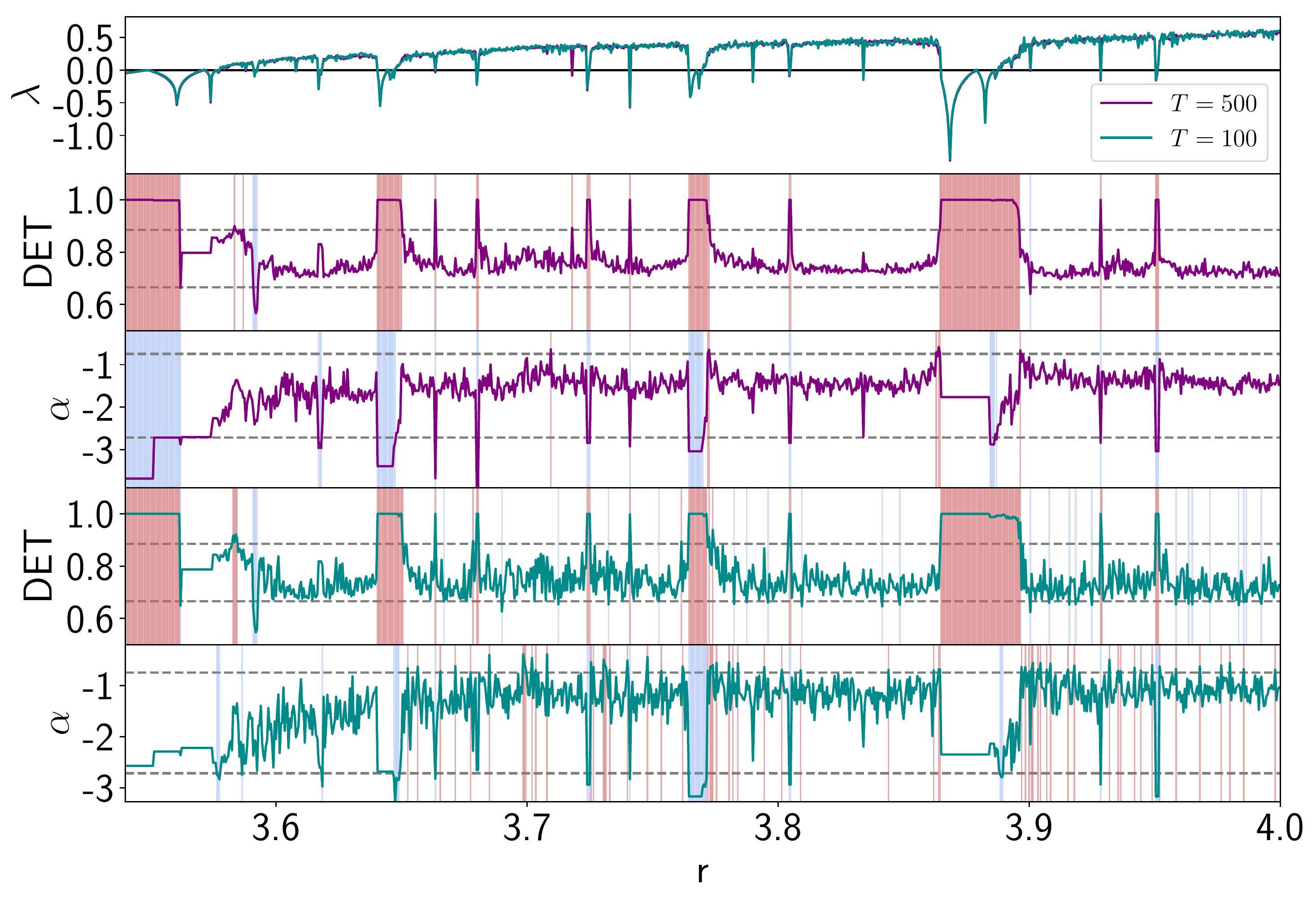}
\caption{$\lambda_1$, DET and the slope $\alpha$ of the RL curve $\alpha$ for varying $r \in [3.5, 3.95]$ with $n=1000$. Different curves correspond to different time series lengths. $95\%\,$-- confidence bounds (gray dashed lines) for both RQA measures are obtained via bootstrapping. Significant transitions outside of these are indicated by blue/red rectangles.}
\label{fig5b}
\end{subfigure}
\caption{Robustness of RL to varying noise intensity and time series length for the logistic map.}
\label{fig5}
\end{figure}

In real--world data sets, several effects can impede the detection of regime transitions. We test the robustness of RL compared to DET as a standard RQA measure for two such effects, i.e. contamination with white noise and short time series lengths.
Figure \ref{fig5a} shows bifurcations for varying noise strength. We plot both DET and the slope $\alpha$ of the RL curve for $r \in [3.5, 3.95]$. RPs are generated as described above but for time series that are contaminated by uncorrelated white noise of different strength. The largest Lyapunov exponent $\lambda_1$ shows that for $\sigma_{\mathrm{n}} = 0.0005$, only few of the chaos-periodic transitions are detected. For $\sigma_{\mathrm{n}} = 0.0001$, some transitions are still well depicted by both measures while others become less prominent already. We evaluate the performance of detecting transitions for $\sigma_{\mathrm{n}} = 0.0005$ by calculating confidence intervals for the noise-free case as a rather strict bound.
Both measures perform with similar success. 
Yet we point out that RL seems to resolve transitions close to strongly periodic dynamics more clearly in presence of noise, e.g. at around $r\approx 3.55$ and $r\approx 3.83$.
Analogously to this analysis, we plot results for different time series lengths in Fig. \ref{fig5a}. As expected, it appears that the number of false detections increases for both measures for shorter time series. However, both still succeed to pinpoint chaos--order transitions even for very short time series with $T=100$. Beyond that, the laminar parameter range is also still identified. Anyway, $\alpha$ indicates more false shifts than $\mathrm{DET}$ which may be due to the rather basic box-counting approach that limits the computation of RL to a few boxes in case of very short time series.

\subsection{Roessler System}
\label{sec3.2}
We further explore the ability of RL to detect regime transitions for continuous dynamical systems with the Roessler system as a standard example (see Tab. 1). We vary $c\in[2,\,10]$ with $n=2000$ different values. Increasingly dominant chaotic behaviour is expected whereas the average distance between unstable periodic orbits decreases for increasing $c$. 
The nonlinearly coupled $x\,$--component is embedded with parameters given in Tab. 1. The vicinity threshold is fixed as $\epsilon = 0.5\sigma_{\mathrm{ts}}$.

\begin{figure}[!h]
\centering
\includegraphics[width=.65\textwidth]{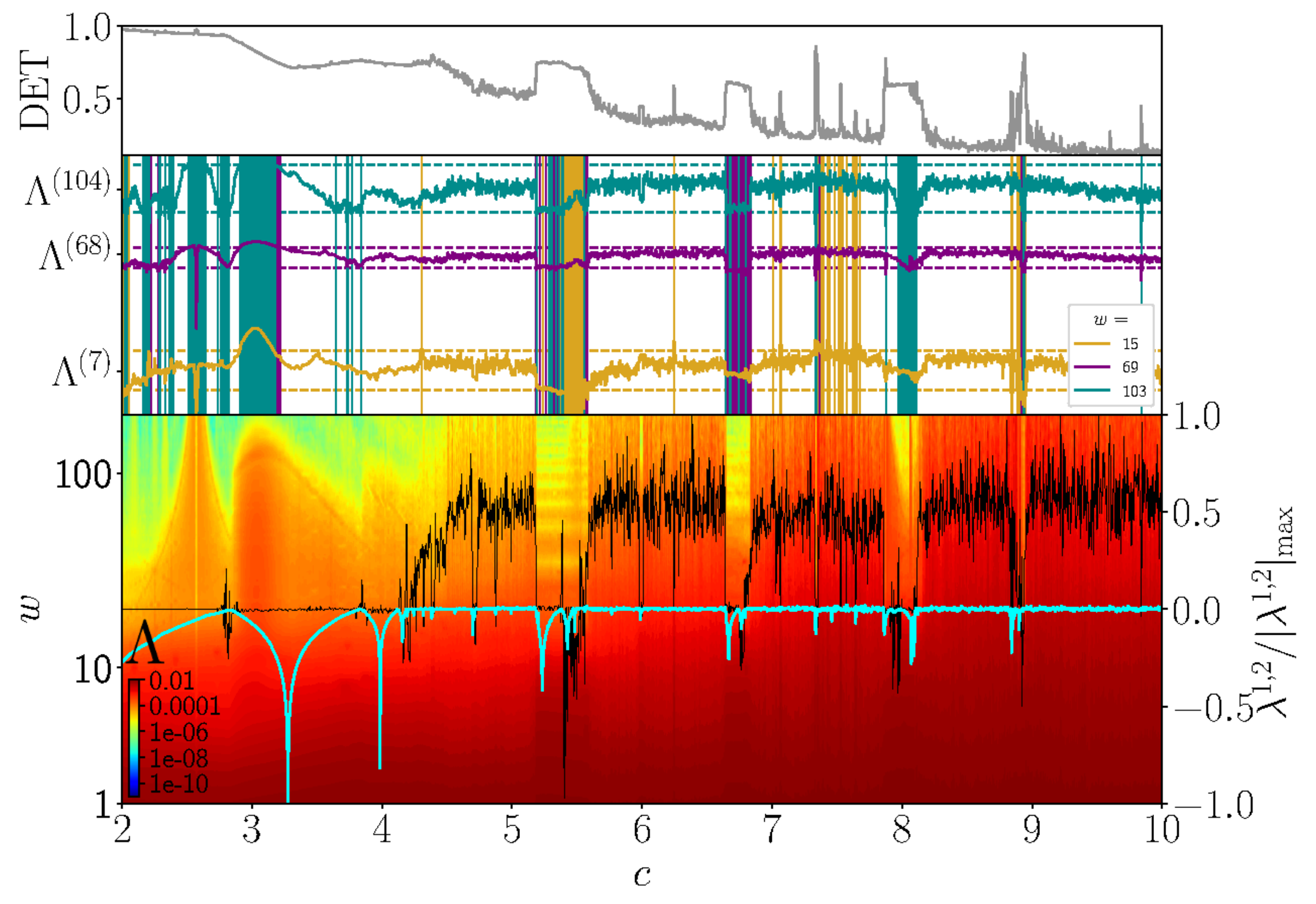}
\caption{$\mathrm{DET}$, single scale-specific lacunarities $\Lambda^{(w)}$ and RL curves for varying $c \in [2, 10]$ with $n=2000$ for the Roessler system. $95\%\,$-- confidence bounds are obtained via boostrapping and excursions outside the indicated horizontals are indicated. Each time series has $T=1000$ values after discarding transients. In the lower panel, $w\,$--axis and color coding are scaled logarithmically. Black (cyan) curves display the first (second) Lyapunov exponent normalized to their maximum absolute values for better visibility.}
\label{fig6}
\end{figure}

\FloatBarrier
Instead of $\alpha$, we analyse single scale-specific values of $\Lambda(w)$ since in general, multiple scaling regions can be identified that are not well represented by a single scaling exponent (see Fig. \ref{fig2}).
In the lower panel of Fig. \ref{fig6} $\lambda_1$ (black) and the second Lyapunov exponent $\lambda_2$ (white) serve as a reference and enable us to localize periodic windows in the regarded parameter range. The color coded RL curves display rich information on various characteristic scales underlying the chaotic and periodic dynamics. 
Note that RL detects bifurcations only captured by both of the displayed Lyapunov exponents. For instance, this can be seen in the range $c\in[2,4]$ where RL corresponds well to $\lambda_2$ for larger box sizes whereas pronounced variations are neither captured by $\lambda_1$ nor by determinism. Above $c=5$, the transitions captured by RL match those indicated by $\mathrm{DET}$. Particularly for the smaller box sizes, a trend in overall complexity for increasing $c$ is present. We subtract this quadratic trend from three lacunarities for fixed box sizes $w$ in the upper panel of of Fig. \ref{fig6} and compute confidence bounds via the introduced bootstrapping procedure. Almost all of the occuring transitions are detected as indicated by the colored vertical lines. Interestingly, certain periodic windows are most prominently captured by distinct box sizes such as at $c\approx 3.7$ or $c \approx 6.4$.
This is due to the fact that as soon as a box size that corresonds to a characteristic scale is reached, the RP becomes more homogenuous on this scale and entails low RL. From this perspective, RL may additionally provide similar spectral information on the time series as usually obtained from Wavelet analysis \cite{aballe1999using} in some cases. 

\FloatBarrier

\subsection{Bistable Noise-Driven System}
\label{sec3.3}
Finally, we demonstrate that RL can also uncover subtle transitions in the time evolution of a nonstationary, stochastic system. To this extent, we study a bistable system which is driven by uncorrelated noise $\xi(t)$ and a periodic  component. Such a system is often illustrated as a Brownian particle trapped in a double-well potential with a periodic driving force \cite{kalmykov2007brownian}.
It can be described by the Langevin equation given in Tab. 1
where $K$ controls the shape of the deterministic potential function, $A$ yields the strength of the periodic component with its frequency $\omega$ and $D$ controls the noise strength. $\xi(t)$ is chosen as Gaussian white noise. Equation \ref{eq4} is solved using the Euler--Maruyama method with a sampling interval of $\Delta t = 0.002$ and subsequent downsampling to $\Delta t = 0.006$.
The interplay of the noise strength and the periodic force
determines the transition dynamics between the two stable fixed points as described by the general phenomenon of stochastic resonance \cite{stochasticresonance}. For fixed $A$, an optimal $D$ exists such that the signal-to-noise ratio (SNR) is maximized. 

\begin{figure}[!h]
\centering
\includegraphics[width=.65\textwidth]{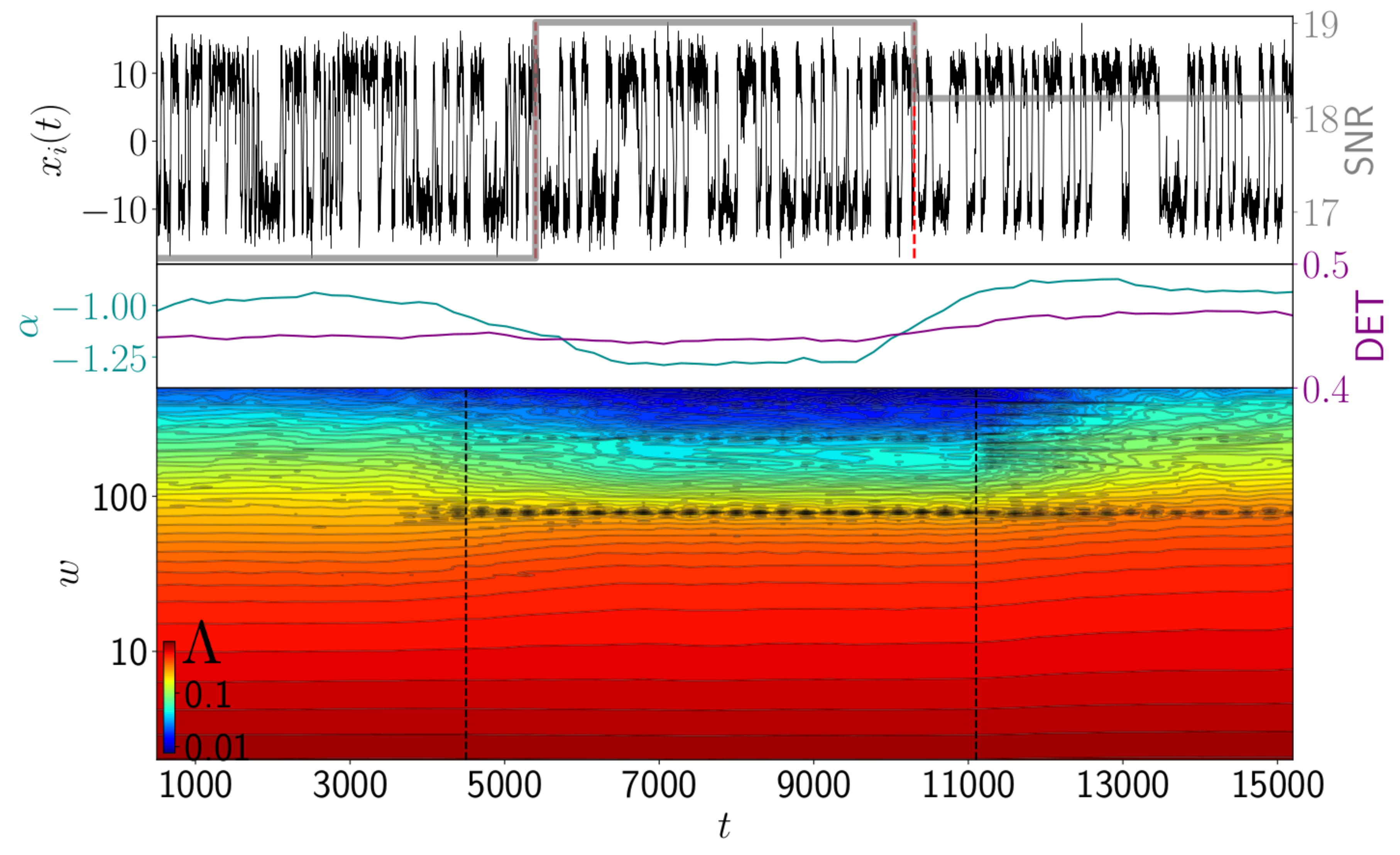}
\caption{Single time series realization and averaged RL curves on sliding windows for bistable noise-driven system. Sample averaged SNR (gray) is indicated in the upper panel, sample averaged $\mathrm{DET}$ (purple) and $\alpha$ (cyan) are compared in the center panel. 
Vertical dashed lines mark the two transitions $\mathcal{X}_1 \rightarrow \mathcal{X}_2$ and $\mathcal{X}_2 \rightarrow \mathcal{X}_3$. For the RL curves, $w\,$--axis and color coding are scaled logarithmically.}
\label{fig7}
\end{figure}

There are several notions of how different regime shifts can be classified \cite{ashwin2012tipping}. A basic differentiation may be made between transitions which are induced by a stochastic or a deterministic component of the respective system. We vary both the frequency $\omega$ of the driving force and noise strength $D$ to test whether RL is able to detect these two different types of transitions.
Both increasing $\omega$ and decreasing $D$ result in a lower SNR since it triggers more high-frequency variability and impedes regular switching.
We generate time series of total length $T = 1.5\cdot 10^{4}$ whereas the frequency $\omega$ is first abruptly decreased from $\omega=20$ to $\omega=8$ at $t_1=T/3$ while noise strength remains constant at $D=36$. At $t_2=2T/3$, noise strength $\sigma$ is decreased from $D=36$ to $D=28$ while retaining $\omega=8$. $100$ realizations $x_i(t)$ with random initial conditions are generated. The height of the potential barrier is $h = K^2/4 = 2500 \gg A$, therefore jumps between the potential minima can not be solely caused by the periodic forcing. However, both noise strengths can result in purely noise-driven jumps. RL is computed for RPs with a fixed recurrence rate of $10\%$ on sliding windows of width $2000$ with a $90\%\,$ overlap for each sample.
In total, the following two shifts are studied:
\begin{align*}
&\mathcal{X}_1 \rightarrow \mathcal{X}_2 \rightarrow \mathcal{X}_3|_{K=100, A=320}: \\ 
& \left(\omega=20, D=36 \right) \xrightarrow{t_1} \left(\omega=8,  D=36\right) \xrightarrow{t_2} \left(\omega=8,D=28 \right)
\end{align*}
between states $\mathcal{X}_{1},\,\mathcal{X}_{2}$ and $\mathcal{X}_{3}$.
The upper panel of Fig. \ref{fig7} shows an example of a sampled time series. We can observe the subtlety of the two parameter shifts which can not be solely localized through visual inspection. In the lower panel, the variation of color-coded RL curves is displayed whereas the average over all generated samples is shown. As expected, the transitions are most striking for the largest considered box sizes $w$ since these have the same order of magnitude as the switching time scales between the two fixed points. Such switches are encoded as gaps in the RPs and can thus be well identified by RL. The decrease of $\omega$ after $\mathcal{X}_1 \rightarrow \mathcal{X}_2$ results in more homogenuous RPs on these time scales since the system on average resides for shorter time periods within one potential minimum. Within the $\mathcal{X}_2$ regime, SNR is enhanced yielding a more regular switching behaviour. This regularity further stands out as an almost stable switching period as indicated by RL at $w\approx 100$. Recurrences within this regime can consequently be regarded as less complex for time periods exceeding this switching cycle.
A similar increase in large-scale RL can be observed for the decrease in noise intensity $\mathcal{X}_2 \rightarrow \mathcal{X}_3$ : as noise-driven transitions are less likely, the RPs are more `gappy' on longer time scales. 
\FloatBarrier
Both states $\mathcal{X}_1$ and $\mathcal{X}_3$ thus share the feature that lower frequency variability is more complex in terms of more diverse recurrent patterns. Anyway, they differ with respect to their high--frequency variability. For small box sizes (mostly reflecting intra-well dynamics), RL is higher in the $\mathcal{X}_3$ regime as indicated by the gray contours. 

When both the driving frequency and noise intensity are low, recurrences are less eratically clustered, yielding more coherent, variable recurrent periods for short durations.
The semi-stable period is preserved in $\mathcal{X}_3$ but less pronounced than in $\mathcal{X}_2$ since lower noise strength results in a weakened SNR through the mechanism of stochastic resonance. 
Finally, we compute three regime-specific RL curves for each of the 100 samples and calculate the sample averaged slopes $\alpha$ for the different regimes. With $\alpha_{\mathcal{X}_1} = -1.09 \pm 0.17\,$, $\alpha_{\mathcal{X}_2} = -1.38 \pm 0.18 \,$ and  $\alpha_{\mathcal{X}_3} = -1.00 \pm 0.18\,$, the two regimes $\mathcal{X}_1$ and $\mathcal{X}_2$ can be well distinguished from $\mathcal{X}_2$ but are similar to each other in the range of their respective standard errors. 
Anyway, the middle panel in Fig. \ref{fig7} shows that $\alpha$ still captures both transitions more clearly compared to $\mathrm{DET}$. While $\mathrm{DET}$ performed superior for short time series (see \ref{fig5b}), RL gives more convincing results for these rather subtle shifts. 

\section{Application to Thermoacoustic Instability Time Series}
\label{sec4}
In order to evaluate the performance of RL as a complexity indicator for empirical data, we apply it to acoustic pressure time series from a laboratory combustor with turbulent flow, operating at atmospheric pressure. The univariate time series was measured with a $10\mathrm{kHz}$ resolution and is known to undergo a rich variety of transitions between chaotic, intermittent and periodic dynamics \cite{godavarthi2017recurrence}. Practical relevance arises e.g. from the use of gas turbine engines for propulsion and power generation. The two main acting subsystems are the unsteady heat release and the acoustic field. Positive feedbacks between both may lead into a state called thermoacoustic instability, enhancing heat transfer to the walls of the combustion chamber and resulting in increased mechanical stress. This can cause severe damages to an aircraft's engine or shutdown in the case of a power plant \cite{kasthuri2019dynamical}. The parameter which can drive the turbulent system into an unstable state is the fuel/air ratio. If it falls below a critical threshold, flame blowout can be the result which leads to an abrupt drop of thrust for an aircraft. This raises the need of effective monitoring to detect impending instabilities and thermoacoustic transitions such including thermoacoustic instability and blowout. In the following, we will apply RL analysis to RPs of the system in order to idenfity the different regimes and localize the shifts between them. Several studies have been carried out in this context, classifying dynamical states with fractal measures \cite{unni2015multifractal}, employing complex networks \cite{unni2018emergence, krishnan2019emergence} and RPs \cite{nair2014intermittency, godavarthi2018coupled, godavarthi2017recurrence, sujith2020complex}.
Special emphasis will be given to the distinction of normal operating conditions (combustion noise) and an impending blowout situation. We first introduce the experimental setup in Sect. \ref{sec4.1}. In Sect. \ref{sec4.2}, we investigate regime shifts in the system in terms of varying RL of acoustic pressure time series.
\subsection{Experimental Setup and Data Acquisition}
\label{sec4.1}
The acoustic pressure ($p'$) data used for the study was obtained from a turbulent combustor with a bluff body stabilized flame. The combustor consists of a rectangular chamber (length = 140 cm, cross section= 9 cm$\times$9 cm) that houses a circular disk (bluff body, diameter = 4.7 cm, thickness = 1 cm) aligned along the axis of the combustion chamber. The bluff body produces stagnation points and recirculation zone in the flow where the flame can be stabilized. The flame is unsteady due to the turbulent fluctuations in the flow. The unsteady fluctuations of the flame and the fluctuations in the acoustic field of the combustion chamber is in feedback with each other. At certain conditions, the feed back is positive causing growth of amplitude of periodic acoustic pressure oscillations inside the combustion. The amplitude of oscillations eventually saturates as the losses (damping) due to the nonlinearities in the system increase with the increase in the amplitude of acoustic pressure oscillations. Such pressure oscillations are known as thermoacoustic instability.  

\begin{figure}[!h]
\centering
\includegraphics[width=.65\textwidth]{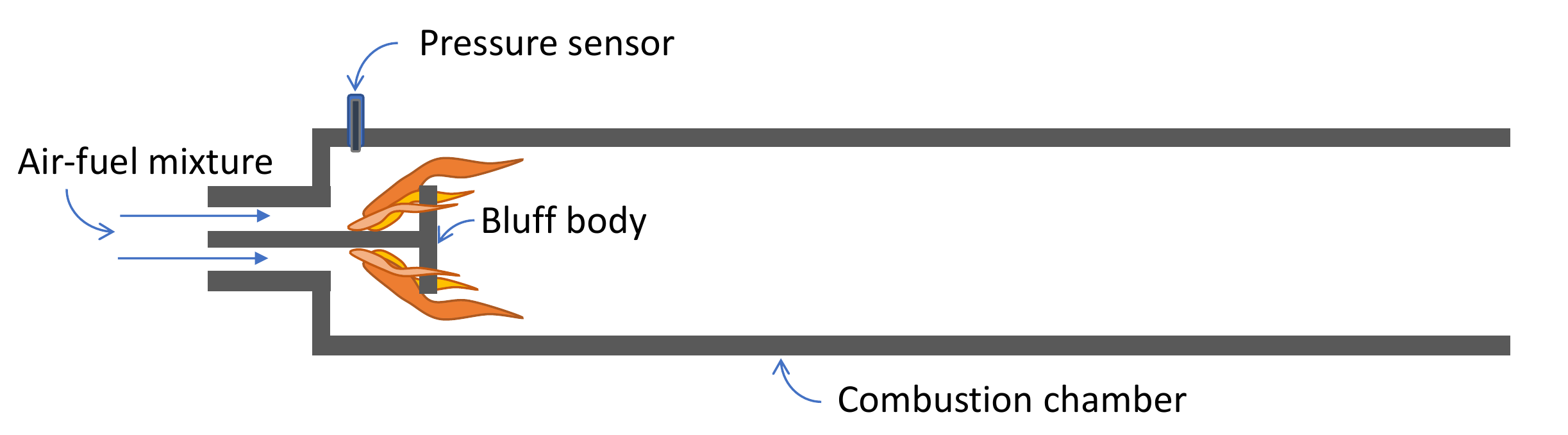}
\caption{Schematic illustration of the combustion chamber employed in the experimental setup.}
\label{fig8}
\end{figure}

The acoustic pressure fluctuations in the combustion chamber is measured using a pressure transducer (PCB103B02) at $1\times 10^4$ samples per second. The fuel used is LPG ($40\%$ Butane, $60\%$ Propane). The flow rate of partially premixed air-fuel mixture is varied in a quasi steady manner by increasing the air flow rate for a fixed fuel flow rate (1.04 g/s). This also reduces the equivalence ratio, $\phi$, of the fuel-air mixture defined as the ratio of actual fuel-air mass flow rate ratio to the stoichiometric fuel-air mass flow rate ratio. As $\phi$ reduces by increasing the airflow rate, initially the pressure fluctuations change from aperiodic oscillations to thermoacoustic instability via intermittency. On further reduction of $\phi$, the pressure oscillations exhibits intermittency post thermoacoustic instability. When $\phi$ is reduced further, we approach flame blowout. Flame blowout is a phenomena where the flame loses its ability to stabilize inside a combustion chamber and hence undergoes extinction. Prior to flame blowout, the pressure oscillations have low amplitude and are aperiodic. Further details of the experimental setup (Fig. \ref{fig8}) and different dynamic regimes are detailed in \cite{unni2015multifractal}. 
\FloatBarrier
\subsection{Results}
\label{sec4.2}
\begin{figure*}[!h]
\centering
\begin{subfigure}[t]{.7\textwidth}
\centering
\includegraphics[width=\textwidth]{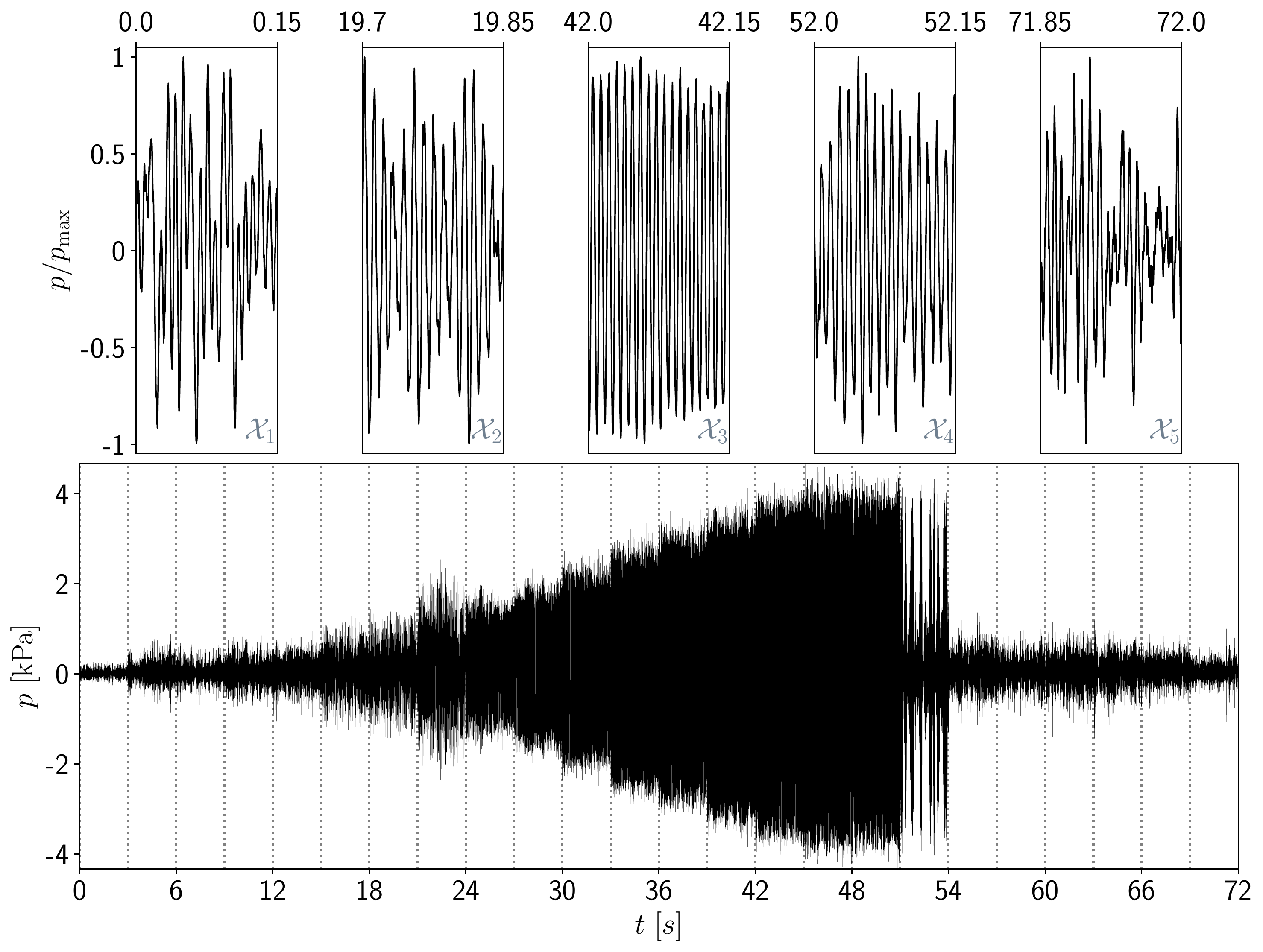}
\caption{Enlarged, normalized time series segments of different dynamical regimes and the entire measurement time series of acoustic pressure in $\mathrm{kPa}$. Each fixed parameter window covers a duration of three seconds. The displayed dynamical states are refered to as combustion noise $(\mathcal{X}_1)$, intermittency prior to instability $(\mathcal{X}_2)$, thermoacoustic instability $(\mathcal{X}_3)$, intermittency after instability $(\mathcal{X}_4)$ and near-blowout oscillations $(\mathcal{X}_5)$ respectively.}
\label{fig9a}
\end{subfigure}
\begin{subfigure}[t]{.7\textwidth}
\centering
\includegraphics[width=\textwidth]{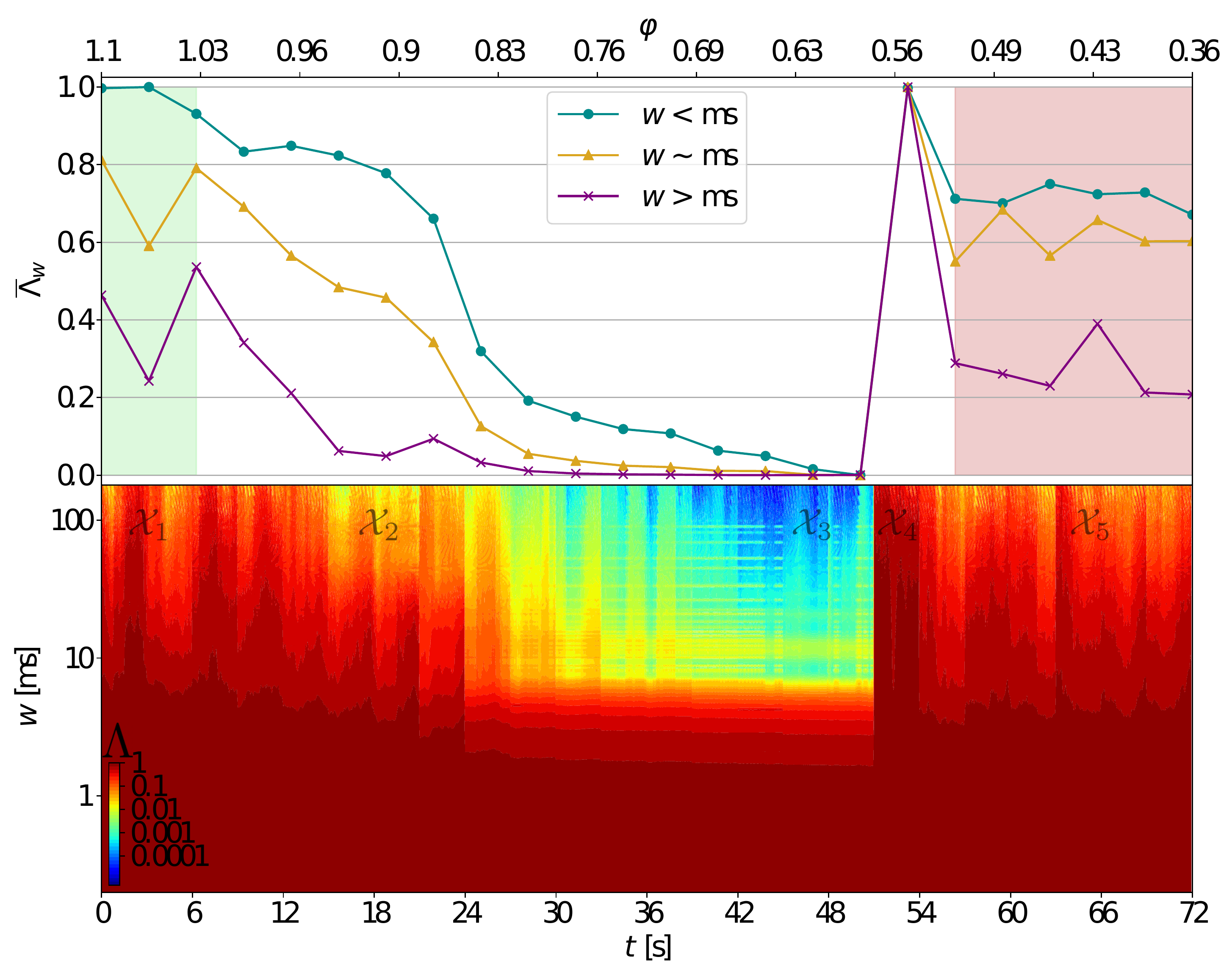}
\caption{Scale-averaged lacunarities $\overline{\Lambda}_w$ and RL curves for full time series computed on sliding windows of $900\,\mathrm{ms}$ width. In the upper panel, green and red shading indicate combustion noise $(\mathcal{X}_1)$ and near blowout oscillations $(\mathcal{X}_5)$ respectively.
Air flow rate is increased in a quasi-static manner after maintaining it steady for a duration of 3 seconds. In the lower panel, $w\,$--axis and color coding are scaled logarithmically.}
\label{fig9b}
\end{subfigure}
\caption{Application of RL to acoustic pressure fluctuation time series from a thermoacoustic combustor.}
\label{fig9}
\end{figure*}

The different operating conditions of a thermoacoustic combustor have been extensively studied in the literature both in laboratory conditions and model system data \cite{kasthuri2019dynamical}. 
Measures based both on RPs and recurrence networks have been successfully applied to detect dynamical transitions between different regimes \cite{godavarthi2018coupled, godavarthi2017recurrence}. 
Prior to our analysis, we can already give a brief classification of the different dynamical states based on these insights. Figure \ref{fig9} shows the full time series in the bottom panel and enlarged segments with normalized amplitude of it in the top panel. The first segment shows combustion noise $(\mathcal{X}_1)$ which is the general term for stable operating conditions. It is comprised of low amplitude aperiodic pressure fluctuations which can be classified as chaotic dynamics. Every dotted gray line in the lower panel marks a decrease in $\phi$. In this sense, each three second sub--time series should be regarded as a separate experiment with constant parameters and will be evaluated as such in the following. As $\phi$ is discontinuously increased along the time axis,
we observe a dynamical state characterized by aperiodic oscillations interrupted by large amplitude harmonic oscillations. This state is generally refered to as intermittency$(\mathcal{X}_2)$ and is observed as a transition state betweeen aperiodic oscillations and thermoacoustic instability.
The third zoomed segment shows thermoacoustic instability $(\mathcal{X}_3)$ which is constituted by periodic large amplitude pressure fluctuations. As $\phi$ is increased further, the periodic oscillations subside and a different state of intermittency $(\mathcal{X}_4)$ is observed which we will refer to as intermittency after instability (in contrast to intermittency prior to instability). The last segment again shows aperiodic oscillations of low amplitude which are a precursor of an impending blowout situation where the flame can not longer be sustained $(\mathcal{X}_5)$.

In order to carry out our analysis, an adequate phase space embedding is required for the measured time series as the underlying system should be regarded as high-dimensional. We apply the mentioned standard methods to fix a suitable embedding delay and dimension. Since we aim at analysing the dynamics of the combustor for a range of parameters, we estimate common embedding parameters appropiate for all different dynamical states. To also evaluate significance of our results for RL, we apply a sliding window analysis to the time series. We choose a window size of $900\,\mathrm{ms}$ with $95\%$ overlap between consecutive windows while no overlap is allowed between the different sub-time series with fixed $\phi$. We ensured that all of the following results are robust in a reasonable range of window and overlap widths. A sufficient tradeoff for all time series segments is obtained by analysing how strongly both embedding parameters fluctuate for the different regimes in time. Embedding delay is maximum for combustion noise and it decreases towards enhanced periodic oscillations. Highest average embedding dimensions are estimated for intermittency prior to instability. We conclude that our global parameter choice of $d = 5 \, , \, \tau = 23$ is suitable for further analysis.

Based on this choice, we compute RPs on sliding windows based on the delay-embedded phase space trajectory of the system. We choose a threshold $\epsilon$ such that it yields constant recurrence rate of $10\%$ for all RPs. All results are qualitatively sustained for reasonable variations of $\epsilon$.
For a visual impression of RPs of a similar system, the reader is pointed to \cite{nair2014intermittency, godavarthi2018coupled}. The procedure is now carried out as follows: we first calculate a RL curve for each obtained RP which refers to a certain time instance for fixed $\phi$. We concatenate the entire set of RL curves to illustrate them in the same fashion as for the synthetic data examples to display variations of complexity on all time scales.
Additionally, we classify the different dynamical regimes by scale-averages of the RL curves.
In order to estimate scale averages of RL, we first average all RL curves for fixed $\phi$ obtained from the sliding window analysis. Next, we average RL values for time scales $w\le 1\,\mathrm{ms}$, $1\,\mathrm{ms}<w<100\,\mathrm{ms}$ and $w\ge 100\,\mathrm{ms}$ separately. Note that these groups cover different numbers of RL values.
The results are shown in Fig. \ref{fig9}.

In the lower panel, we observe a gradual descent of RL at all time scales $w$ from combustion noise into the impending instability, interrupted by spikes of varying amplitude. The overarching trend until about $51\mathrm{s}$ shows a reduction in dynamic complexity from the aperiodic, chaotic oscillations during stable operation into periodic oscillations during thermoacoustic instability. By means of RL, we can interpret this decrease as weakened heterogenity of recurrences, entailing that the system shows temporal patterns with less strong variability as it approaches the instability regime. 
The intermittency route into instability becomes well visible by the shape of the RL curves: the cut-off value of the truncated power law decay is continuously reduced, displaying a significant shift of characteristic time scales in the pressure fluctuations of the combustor. Cut-off values can be infered approximately from the graph by tracking sudden color switches.
In the range between $24-38\mathrm{s}$, intermittency manifests itself in the episodic reduction of large scale heterogenity in the RPs. The system jumps between harmonic and aperiodic oscillations and thus shifts its characteristic time scale discontinuously until multiple coexisting periods (horizontal yellow lines) are reduced to a single dominant period. 
A sharp rise of RL at all scales marks the slow intermittency regime prior to the near blowout situation.
The average level of complexity during this lean blowout state appears close to that observed for combustion noise at the beginning of the measurement series. In the upper panel, the scale-averages yet uncover a difference in sub-ms RL between combustion noise and lean blowout state: the aperiodic oscillations in the former regime seemingly occur in a more heterogenous fashion than for the latter. The two other scales-averages do not indicate a remarkable difference between these two regimes. 
Note that the scale averages are rescaled to $[0,1]$ for better comparison. 
Consequently, RL enables us to detect five distinct regimes and to track respective variations in the characteristic time scale of the system. These findings generally corroborate those from earlier studies. Furthermore, it detects a subtle difference in the complexity between the dynamically similar combustion noise state and the near blow out situation.

\section{Conclusion}
\label{sec5}
We have put forward a novel recurrence quantification measure to quantify the degree of complexity of nonlinear time series, namely recurrence lacunarity.
The identification of different dynamical regimes in multiple real--world applications has attracted a lot of interest in the literature. The method we propose contributes to this toolbox of complexity indicators by representing a time series as a recurrence plot and characterizing its heterogenity on all time scales relevant to the system. Even though both recurrence plots and lacunarity are broadly acknowledged as powerful stand-alone tools, we have demonstrated that their combination can yield valuable insights in the context of regime shift detection. 
The method's main advantage is that it is able to capture multiscale features of the recurrence plot while traditional measures are based on line structures which can always only encode a certain aspect of the dynamics. Our approach does not require specification of minimum line lengths. 
\FloatBarrier
It naturally yields information both on the heterogenity of recurrence plots and their scaling properties, opening up a new perspective of analysing point statistics in partitioned RPs rather than constricted structures.
We have shown that the applicability of lacunarity to general nonlinear dynamical systems by means of the well established recurrence plot approach can be fruitful.

As this work's main focus was on the identification of dynamical regime shifts, we have studied three different paradigmatic systems in detail to showcase the potentials of the method.
We have found that the proposed method is able to uncover transitions of different origin even in presence of noise and for short time series which makes it broadly applicable to many real--world phenomena. 
We have further demonstrated that recurrence lacunarity may also be useful in characterizing subtle transitions of different nature. In comparison to DET as a traditional recurrence quantifier, it showed comparable well performance for noisy time series. Even though the results were less convincing for a short time series, RL performed superior for subtle transitions in the bistable system.

Finally, we have employed a system exhibiting thermoacoustic instability to study whether our approach enables us to detect regime shifts in an (experimental) real world system. Our method has enables us to identify the different dynamical transitions that the system undergoes.
We found that the intermittency route from stable operation into thermoacoustic instability manifests itself as a continuous transition from aperiodic to harmonic oscillations.
We have ultimately addressed the challenge of differentiating between the dynamically similar but practically different states of stable operation and a near blowout situation. It appeared that short-term acoustic pressure fluctuations show less variable temporal recurrent patterns during a near-blowout situation than during regular operating mode. How this can be interpreted and whether it can also be captured in terms of (nonlinear) serial dependence should be addressed in future work.

Furthermore, it appears as a promising direction for future work to investigate in more detail the relations between the fractal scaling of RPs and fractality of the underlying time series and the attractor dimension. Including RL in feature selection approaches may also improve the performance of machine learning techniques that classify nonlinear data \cite{ye2018classification}. Another line of research should be concerned with the robustness of the proposed method against spurious effects introduced by erroneous embedding and RP related pitfalls when compared to traditional RQA measures \cite{kraemer2019border, marwan2011avoid}. 

\section*{Acknowledgements}
This research was supported by the Deutsche Forschungsgemeinschaft in the context of the DFG project MA4759/11-1 `Nonlinear empirical mode analysis of complex systems: Development of general approach and application in climate', the DFG project MA4759/9-1 `Recurrence plot analysis of regime changes in dynamical systems' and it has received funding from the European Union’s Horizon 2020 research and innovation programme under grant agreement No 820970 as a TiPES contribution.
RIS acknowledges the Science and Engineering Research Board (SERB) of the Department of Science and Technology, Government of India for the funding under the grant Nos.: DST/SF/1(EC)/2006 (Swarnajayanti Fellowship) and JCB/2018/000034/SSC (JC Bose Fellowship). VRU thanks University of California San Diego for the postdoctoral fellowship.  

\section*{Conflict of interest}
The authors declare that they have no conflict of interest.

\bibliographystyle{ieeetr}
\bibliography{bibli}  






\end{document}